# Molecular polaritonics: Chemical Dynamics under strong Light-Matter Coupling

Tao E. Li[a], Bingyu Cui[a,b], Joseph E. Subotnik[a] and Abraham Nitzan[a,b]

(*a*) Department of Chemistry, University of Pennsylvania, Philadelphia, PA 19104, USA

(*b*) School of Chemistry, Tel Aviv University, Tel Aviv 69978, Israel

## Abstract

Chemical manifestations of strong light-matter coupling have been recently subject of intense experimental and theoretical studies. Here we review the present status of this field. Section 1 is an introduction to molecular polaritonics and to collective response aspects of light-matter interactions. Section 2 provides an overview of the key experimental observations of these effects while Section 3 describe our current theoretical understanding of the effect of strong light-matter coupling on chemical dynamics. A brief outline of applications to energy conversion processes in given in Section 4. Pending technical issues in the constructions of theoretical approach are briefly described in Section 5. The summary in Section 6 also outlines the paths ahead in this exciting endeavor.

## 1. Introduction and background

The response of chemical systems to light, a central theme in chemistry research and applications, has been studied along two main lines: spectroscopy and photochemistry. In both, a standard assumption is that the incident field is not affected by the presence of the molecule or, apart from the homogeneous environmental dielectric response, its immediate vicinity. More recently, attention has focused on circumstances where the structure of the immediate molecular environment does modify the local electromagnetic (EM) field, so that the field experienced by the molecule may be substantially different from the incident (far) field. The underlying physics of surface-enhanced Raman scattering is an important example, in which the field experienced by a molecule as well as the resulting molecular behavior are dominated by the geometry and the plasmonic response of a nearby metallic structure. More generally, the molecular consequences of light-matter coupling can strongly depend on the presence of an inhomogeneous dielectric environment. Of particular interest are situations where the structure and dielectric properties of such environments can be controlled. Grated metal surfaces, arrays of metal particles, and optical cavities – Fabry-Pérot (FP) cavities where the molecular system is confined between two mirrors and plasmonic cavities, where molecules are located in nanogaps between metal particles, are examples of current interest.

An important consequence of quantum electrodynamics (QED) is that inside a cavity or in the proximity of a metal structure, both the strength of molecule-field interaction and the density of EM field modes can be altered. In the simplest case of a cavity enclosed by perfectly reflecting boundaries the light-matter interaction and the density of field modes scale like $\mathcal{V}^{-1/2}$ and $\mathcal{V}$, respectively, where $\mathcal{V}$ is the cavity volume. In practice, metallic boundaries are never perfectly reflecting, and damping of field oscillations inside the metal can become the dominant loss mechanism for molecular photo-processes in such environments. Sometimes the combined effects of this damping and other losses in the molecular subsystem can mask the effect of strong molecule-field coupling in such a system. We refer to situations where the effect strong coupling is observable despite any damping as the "strong coupling limit". Simple examples are provided by two linearly coupled damped harmonic oscillators(1) or by the quantum dynamics of two coupled two-level species described as follows: Denote the lower and upper states of species C and M (henceforth referred to as "cavity mode" and "molecule") by $(g_c, x_c)$ and $(g_0, x_0)$ with corresponding level spacings $\omega_c$ and $\omega_0$, and consider the ground $|G\rangle = |g_c\rangle|g_0\rangle$ and two lower excited states $|C\rangle = |x_c\rangle|g_0\rangle$ and $|X\rangle = |g_c\rangle|x_0\rangle$ of the global system. These states are energetically close to each other if $\omega_c \simeq \omega_0$, and in the ensuing dynamics the highest excited state $|x_c, x_0\rangle$ is disregarded. Setting the global ground state energy to zero, the Hamiltonian is

$$\hat{H} = \hbar\omega_c|C\rangle\langle C| + \hbar\omega_0|X\rangle\langle X| + \hbar g_0|C\rangle\langle X| + \hbar g_0^*|X\rangle\langle C| = \hbar\omega_+|+\rangle\langle +| + \hbar\omega_-|-\rangle\langle -|, \quad (1)$$

where $g_0$ is the molecule-cavity mode coupling and the eigenvalues $\omega_\pm$ and their separation $\Delta\omega_\pm$ are given by

$$\omega_\pm = (1/2)\left[\omega_c + \omega_0 \pm \sqrt{(\omega_c - \omega_0)^2 + 4|g_0|^2}\right]; \quad \Delta\omega_\pm = \sqrt{(\omega_c - \omega_0)^2 + 4|g_0|^2} \quad (2)$$

The minimum separation

$$\Omega_R = 2|g_0| \quad (3)$$

is the Rabi frequency for this model. Next, assume that C and M are coupled to their own dissipative environment so that when $g_0 = 0$ their amplitudes evolve according to $\exp\left(-i\omega_j t - (1/2)\gamma_j t\right)$, $j = c, 0$. The ground state absorption lineshape for this system is given by(2)

$$L(\omega) \sim \mathrm{Im}\frac{F(\omega)}{(\omega - \tilde{\omega}_+)(\omega - \tilde{\omega}_-)}, \quad (4)$$

where $\tilde{\omega}_\pm$ are obtained from $\omega_\pm$ by replacing in Eq. (2) $\omega_c$ and $\omega_0$ by $\tilde{\omega}_c = \omega_c - (i/2)\gamma_c$ and $\tilde{\omega}_0 = \omega_0 - (i/2)\gamma_0$, respectively, while $F(\omega) \equiv |\mu_c|^2(\omega - \tilde{\omega}_c) + |\mu_0|^2(\omega - \tilde{\omega}_0) - \mu_c g_0 \mu_0^* - \mu_0 g_0^* \mu_c^*$. Here $\mu_j = \langle g_j|\hat{\mu}_j|x_j\rangle$, $j = c, 0$, is the ground-to-excited transition dipole elements of the two systems. $F(\omega)$ is a relatively weak function of ω, so the structure of the lineshape (4) is determined by the denominator $\left[(\omega - \tilde{\omega}_+)(\omega - \tilde{\omega}_-)\right]^{-1}$. For $\gamma_c = \gamma_0 = \gamma$, the condition for this function to show two peaks is(2) $\Delta\omega_{\pm} > \gamma$, and for the distance of their closest approach, where $\omega_c = \omega_0$, this yields

$$\Omega_R > \gamma \tag{5}$$

Eq. (5) defines the strong coupling limit for this model. Note that "strong coupling" is measured not by the coupling magnitude but by the observability of its consequence.

An important aspect of this strong coupling phenomenon is its collective property. Consider an extension of the model (1) in which *N* identical two-level molecules whose states are labeled $\left(g_j, x_j\right)$, $j = 1,...,N$, interact with the cavity mode C but not with each other. Limiting ourselves again to the single excitation subspace above the zero-energy ground state, the Hamiltonian is

$$\hat{H} = \hbar\omega_c \left|C\right\rangle\left\langle C\right| + \sum_{j=1}^{N}\left(\hbar\omega_0 \left|X_j\right\rangle\left\langle X_j\right| + \hbar g_0 \left|C\right\rangle\left\langle X_j\right| + \hbar g_0^* \left|X_j\right\rangle\left\langle C\right|\right) \tag{6}$$

where now $\left|C\right\rangle = \left|x_c\right\rangle \prod_{j=1}^{N}\left|g_j\right\rangle$ and $\left|X_j\right\rangle = \left|g_c\right\rangle\left|x_j\right\rangle \prod_{l\neq j}\left|g_l\right\rangle$. The consequence of coupling in the Hamiltonian (6) are easily realized by linearly transforming the set of states $\left\{\left|X_j\right\rangle\right\}$ to a new orthonormal set

$$\left|B\right\rangle = \frac{1}{\sqrt{N}}\sum_{j=1}^{N}\left|X_j\right\rangle \; ; \qquad \left|D_l\right\rangle = \sum_{j=1}^{N} d_{lj}\left|X_j\right\rangle; \;\; l = 1,...,N-1 \tag{7}$$

Since $\sum_{j=1}^{N} d_{lj} \sim \left\langle B \mid D_l\right\rangle = 0$ for all *l*, none of the states $\left|D_l\right\rangle$ is coupled to $\left|C\right\rangle$. Furthermore, because $\left\langle G \left| \sum_{j=1}^{N}\hat{\mu}_j \right| D_l \right\rangle = \mu_0 \sum_j d_{lj}$, none of these states is coupled radiatively to the ground state. These so-called *dark states* are indeed dark with respect to the radiative coupling with the ground states as well as the coupling to system C. Furthermore, the coupling between the cavity mode and the so-called bright state B is given by

$$\left\langle C\left|\hat{H}\right|B\right\rangle = \sqrt{N}g_0 \tag{8}$$

Similarly, B is the only excited state of the molecular ensemble that can be reached from the ground state, with the radiative coupling

$$\left\langle G\left|\sum_j \hat{\mu}_j\right|B\right\rangle = \sqrt{N}\mu_0 \tag{9}$$

We are therefore back to the model of Eq. (1) however with a collective coupling $\sqrt{N}g_0$ replacing $g_0$, and with the absorption lineshape, Eq. (4), multiplied by *N*. The latter observation is not unexpected: The integrated absorption of *N* molecules should scale like *N*. However, significantly, the Rabi splitting, Eq. (3), now becomes

$$\Omega_R(N) = 2\sqrt{N}|g_0| \tag{10}$$

showing a remarkable signature of collective response and indicating that strong coupling may be achieved by increasing the number of interacting molecules. Indeed, the observation of split-peaks with separation (Rabi splitting) scaling like the square root of the molecular density has been the hallmark of many recent observations of strong light-matter coupling.

Another expression of collective behavior is manifested in the spontaneous radiative decay of the bright state. Eq. (9) together with the golden rule expression for the radiative decay rate imply that

$$\Gamma_R(N) = N\Gamma_R \tag{11}$$

where $\Gamma_R = \Gamma_R(1)$ is the radiative decay rate of a single molecule. If, in addition to radiative decay, each molecule undergoes a non-radiative relaxation to its own relaxation channel with a rate $\Gamma_{NR}$, the bright state relaxation will appear as

$$\left|\left\langle B(t) \mid B(t=0)\right\rangle\right|^2 = \left(\frac{1}{N}\sum_j \left\langle X_j \mid X_j\right\rangle\right)^2 e^{-\left[\Gamma_{NR}+\Gamma_R(N)\right]t} = \exp\left[\left(-\Gamma_{NR} - N\Gamma_R\right)t\right] \tag{12}$$

In the absence of other relaxation processes, the radiative quantum yield is given by

$$Y_R = N\Gamma_R \int_0^\infty dt \left|\langle B(t) \mid B(t=0)\rangle\right|^2 = \frac{N\Gamma_R}{\Gamma_{NR} + N\Gamma_R} \tag{13}$$

and monotonously increases with *N*.

The simple model described above is a prototype of phenomena encountered when a group of molecules interacts collectively with the radiation field in an optical cavity or in proximity to suitable plasmonic structures, provided that the size of the molecular system is much smaller than the wavelength of the energetically relevant radiation field modes. The two-level system C represents a cavity mode or a plasmonic resonance, which is more commonly modeled as a harmonic oscillator (see Eq. (16) below). The other *N* two-level systems represent *N* molecules, and their coupling to the cavity mode is determined by the molecular transition dipole $\mu_0$ and by the cavity volume $\mathcal{V}$

$$g_0 = \mu_0 \sqrt{\frac{\omega_c}{2\hbar\epsilon_0 \mathcal{V}}}. \tag{14}$$

where $\epsilon_0$ is the vacuum permeability. Light-matter interaction in such a system is manifested by the appearance of the two states $|+\rangle$ and $|-\rangle$, referred to as the upper and lower polaritons (UP and LP), which are hybrid light-matter states. Moreover, these are collective states of the *N*-molecule systems. This by itself is not a unique manifestation of the cavity environment. The bright state $|B\rangle$ is by itself a collective eigenstate of the molecular subsystem, regardless of its coupling to the cavity mode, and its presence is manifested in the ensuing radiative relaxation, Eq. (11), see, e.g. (3). Out of cavity, this state is energetically embedded within the large group of dark states and can be quickly destroyed by other interactions such as intermolecular collisions, not included in Eq. (6). In the cavity, the shifted polariton states are largely out of the dark-states band and are expected to live longer. This is indeed often observed under strong coupling, in particular for the lower polariton.

The models and processes discussed above are simplified versions of more general models that describe some of the rich scopes of phenomena associated with light-matter interaction. The Dicke Hamiltonian,(4; 5) a model for a system of two-level molecules interacting with the transverse radiation field

$$\hat{H}_{coll} = \hbar\sum_{j=1}^{N}\omega_j\hat{\sigma}_j^+\hat{\sigma}_j^- + \sum_q \hbar\omega_q\hat{a}_q^+\hat{a}_q - \hbar\sum_{j=1}^{N}\sum_q g_{qj}(e^{i\mathbf{k}_q\cdot\mathbf{r}_j}\hat{a}_q + e^{-i\mathbf{k}_q\cdot\mathbf{r}_j}\hat{a}_q^\dagger)\hat{\sigma}_{jx} \tag{15}$$

with $g_{qj} = (\omega_q / \epsilon_0\hbar\mathcal{V})^{1/2}\mathbf{e}_q\cdot\hat{\boldsymbol{\mu}}_j$, $\hat{\sigma}_j^+ = |x_j\rangle\langle g_j|$, $\hat{\sigma}_j^- = |g_j\rangle\langle x_j|$ and $\hat{\sigma}_{jx} = \hat{\sigma}_j^+ + \sigma_j^-$ has been used, with $\omega_j = \omega_0$ the same for all molecules, to describe collective spontaneous emission from a system of excited two-level atoms, as well as the statistical properties of the emitted light. It has been observed in different systems, including molecular aggregates. (6)

A simpler version of the Dicke Hamiltonian, setting $\mathbf{r}_j = 0$ for all $j$ when the atomic system size is much smaller than $2\pi\omega_0/c$ (c = speed of light), taking a single radiation field mode to represent a cavity mode, and assuming a uniform coupling $g_{qj} = g_0$ - the same for all molecules, was used by Tavis and Cummings (TC)(7; 8) to describe the main physical consequences of light-matter interaction in an optical cavity. The TC Hamiltonian

$$\hat{H}_{TC} = \hbar\omega_c\hat{a}^\dagger\hat{a} + \hbar\omega_0\sum_{j=1}^{N}\hat{\sigma}_j^+\hat{\sigma}_j^- + \hbar g_0\sum_{j=1}^{N}\left(\hat{a}^\dagger\hat{\sigma}_j^- + \hat{a}\hat{\sigma}_j^+\right) \tag{16)a$$

also makes, in the TC original work, the rotating wave approximation by disregarding terms of the type $\hat{a}\hat{\sigma}_j^-$ and $\hat{a}^\dagger\hat{\sigma}_j^+$ that cannot conserve energy at the lowest order of the interaction. Eq. (16)a is an extension to many atoms of the well-known Jaynes-Cummings model(9) (10) – a two-level system interacting with one harmonic oscillator. If we further approximate each two-level system as a harmonic oscillator (a valid approximation provided that the system is only weakly excited) and replace $\hat{\sigma}_j^+$ $(\hat{\sigma}_j^-)$ in Eq. (16) by bosonic raising and lowering operators $\hat{b}_j^\dagger$ $(\hat{b}_j)$, the TC Hamiltonian becomes

$$\hat{H}_{TC} = \hbar\omega_c\hat{a}^\dagger\hat{a} + \hbar\omega_0\sum_{j=1}^{N}\hat{b}_j^\dagger\hat{b}_j + \hbar g_0\sum_{j=1}^{N}\left(\hat{a}^\dagger\hat{b}_j + \hat{a}\hat{b}_j^\dagger\right) \tag{16)b$$

In analogy to Eq. (7) we can define the raising and lowering operators of the bright mode and the $N-1$ dark modes

$$\hat{B}^{+}=\frac{1}{\sqrt{N}}\sum_{j=1}^{N}\hat{b}_j^{\dagger}\ ;\ \hat{B}^{-}=\frac{1}{\sqrt{N}}\sum_{j=1}^{N}\hat{b}_j \tag{17)a$$

$$D_k^{\dagger}=\sum_{j=1}^{N}d_{kj}^{*}\hat{b}_j^{\dagger}\ ;\ \hat{D}_k^{-}=\sum_{j=1}^{N}d_{kj}\hat{b}_j \qquad ;\ d_{kj}=\frac{1}{\sqrt{N}}\exp\left(-\frac{2\pi ik}{N}j\right) \tag{17)b$$

The transformed Hamiltonian then takes the form

$$\hat{H}_{TC}=\hbar\omega_c\hat{a}^{\dagger}\hat{a}+\hbar\omega_0\hat{B}^{+}\hat{B}^{-}+\frac{1}{2}\hbar\Omega_N\left(\hat{a}^{\dagger}\hat{B}^{-}+\hat{a}\hat{B}^{+}\right)+\hat{H}_D, \tag{18}$$

where $\Omega_N=\Omega_R\left(N\right)$, given by Eq. (10), is the (collective) Rabi frequency, while $\hat{H}_{\mathrm{D}}=\hbar\omega_0\sum_{k=1}^{N-1}\hat{D}_k^{+}\hat{D}_k^{-}$ is the Hamiltonian for the dark modes. The coupled system described by the first three terms of (18) is reminiscent of the Jaynes-Cumming model(9; 10), except that $\hat{B}$ represents an $N$- level system. Proceeding to diagonalize this interacting subspace we get

$$\hat{H}_{TC}=\hbar\omega_{+}\hat{P}_{+}^{\dagger}\hat{P}_{+}+\hbar\omega_{-}\hat{P}_{-}^{\dagger}\hat{P}_{-}+\hat{H}_D. \tag{19}$$

where $\hat{P}_{\pm}^{\dagger}=X_{\pm}^{(\mathrm{B})}\hat{B}^{+}+X_{\pm}^{(\mathrm{C})}\hat{a}^{\dagger}$ with $X_{+}^{(\mathrm{B})}=-X_{-}^{(\mathrm{C})}=\sin\left(\theta\right)$, $X_{-}^{(\mathrm{B})}=X_{+}^{(\mathrm{c})}=\cos\left(\theta\right)$ and $\theta=\frac{1}{2}\tan^{-1}\left(\frac{\Omega_N}{\omega_c-\omega_0}\right)$, and where

$$\omega_{\pm}=\frac{1}{2}\left[\omega_0+\omega_c\pm\sqrt{\Omega_N^2+(\omega_0-\omega_c)^2}\right]. \tag{20}$$

The operators $\hat{P}_{\pm}^{\dagger}$ and $\hat{P}_{\pm}$ create and destroy the upper (+) and lower (-) polaritons – here hybrids between light and molecular electronic excitations that are referred to as exciton-polaritons. The smallest separation between their frequencies, obtained when $\omega_c=\omega_0$, is the Rabi frequency (or Rabi splitting) $\Omega_N$. The analogy to the formulation in Eqs. (1)-(10) is obvious. In fact, like the Hamiltonian (6), the more general Hamiltonian (16) or (18) represents subsystems with different numbers of total excitation, and because it conserves this number its matrix representation is block-diagonal with each block corresponding to a different integer number of excitations. Eqs. (1)-(10) correspond to the single excitation block of this Hamiltonian. These blocks will be coupled by any perturbation (such as an external radiation field) that can change the number of excitations in the system and the system dynamics may further be affected by relaxation processes, giving rise to observables that interrogate the combined effect of relaxation and polaritonic level structure, as exemplified by the lineshape function (4).

To address realistic situations of chemical systems, the above models should be endowed with specific input of the molecular systems under study. Even on a generic level, a realistic model should include the following elements: (a) The nature of the confined electromagnetic environment. Broadly, we distinguish between FP type cavities that can be modeled considering a molecular system confined between two (partially) reflecting mirrors, in which properties of the EM field and its coupling to matter are derived from geometry, and plasmonic cavities where the plasmonic response of the metal also plays an important role. Furthermore, more than one mode supported by these cavities can be relevant to the process under study. Finally, if the size of the molecular system is similar to or larger than the wavelength of relevant cavity modes, the spatial dependence of their interaction must be taken into account. (b) In an anisotropic cavity environment, the molecule-field interaction depends on the molecular orientation, naturally giving rise to a distribution of coupling parameters, e.g. $g_0 \rightarrow g_j$ in Eq. (16) and similarly $\mu_0 \rightarrow \mu_j$ for the radiative coupling to the ground state. (c) While atoms can be often modeled as two-state systems, molecules are way more complex. When considering strong coupling between molecular electronic transitions and cavity modes, the effect of nuclear motion should be taken into account. At the very least, the molecular term in Eq. (16) should be rewritten as $\hbar\omega_0 \sum_{j=1}^{N} \omega_j\left(R_j\right)\hat{\sigma}_j^+\hat{\sigma}_j^- + \hbar\sum_{j=1}^{N} g_j\left(R_j\right)\left(\hat{a}^\dagger\hat{\sigma}_j^- + \hat{a}\hat{\sigma}_j^+\right)$, where $R_j$ represents the nuclear coordinates of molecule *j*. However, this generalization of Eq. (16) relies on the Born-Oppenheimer approximation and should be regarded with caution.(11; 12). In recent years, interest has focused on vibrational strong coupling (VSC),(13; 14) (15) (16) (17) (18) (19-26) where the cavity mode is close to resonance with, and may alter the dynamics of, molecular nuclear motions. (d) Intermolecular interactions should be taken into account. A (conceptually) simple extension of the model (6) is to add a term of the form $\sum_j \sum_{j'\neq j} J_{jj'} \left|X_j\right\rangle\left\langle X_{j'}\right|$ to model energy transfer between excited molecules. More generally, an extended form of Eq. (16)

$$\hat{H} = \sum_{j=1}^{N} \hbar\omega_j \hat{\sigma}_j^+\hat{\sigma}_j^- + \sum_{i\neq j=1}^{N} \hbar J_{ij}\hat{\sigma}_i^+\hat{\sigma}_j^- + \hbar\omega_c \hat{a}^\dagger\hat{a} + \sum_{j=1}^{N} \hbar g_j\left(\hat{a}^\dagger\hat{\sigma}_j^- + \hat{a}\hat{\sigma}_j^+\right) \qquad (21)$$

which accounts for the possibilities of different site energies $\omega_j$ and different molecule-cavity mode interactions $g_j$ and includes a term that describes inter-site energy transfer, has been used (see, e.g., (27)) to described the effect of strong coupling to a cavity mode (last term in Eq. (21)) on exciton transport (itself modeled by the first two terms of (21)). (e) The simple relaxation model used above (see text above Eq. (4)) should be enhanced to include dephasing and thermal effects. (f) Attention should be given to the validity of the description of light-matter interaction. The rotating wave approximation used in Eq. (16) fails in the ultra-strong coupling limit, (28) when $\Omega_N$ is of the order of or larger than the molecular frequency $\omega_0$. Moreover, by Eq. (20), for large enough $\Omega_N$ the lower polariton frequency can be (unphysically) negative, indicating failure of the TC Hamiltonian in the ultrastrong coupling limit. The latter issue is related to the proper Hamiltonian formulation of coupled light-matter systems in the strong interaction limit(29) (30) (31) (32) (33) (34) (35) (28) (36) (37) which is briefly discussed in Section 5.

From the chemistry perspective, “polariton chemistry” offers some interesting possibilities. First, as just discussed, the potentially controlled energy shift of optically active states suggests a route for photochemical control. Second, the stabilization of collective states raises the possibility to harness collective response to affect chemical kinetics. Third, the mixing of molecular states by their mutual coupling to the local radiation field can potentially change the nuclear potential energy surfaces that determine chemical configurations and dynamics. Understanding the way molecules behave under strong light-matter coupling is a prerequisite to advancing these possibilities. This review describes the present status of experimental and theoretical efforts in this direction, focusing mostly on phenomena in which collective behavior is manifested. This will supplement several other reviews of this subject that were published in the last few years(38-41) (42) (43), as well as the review of Owrutsky et al that appears in the present volume.

## 2. Experimental realizations of Strong coupling

Strong EM field-matter coupling is usually achieved by confining the molecular system to small cavities, as implied by Eq. (14), or by using plasmonic nanostructures for field localization and enhancement. A standard setup of the first kind is the FP cavity where the molecular system is confined between a pair of (ideally infinite) parallel planar mirrors separated by a distance $L_c$. A cavity mode is characterized by its quantized, $k_\perp = m\pi/L_c$ , $m = 0, \pm 1 \ldots$, component of the **k** vector in the direction normal to the mirrors, and the 2-dimensional parallel component $\mathbf{k}_{||}$, with the dispersion relation

$$\omega_c\left(k_{||}\right) = \frac{c}{n}\sqrt{k_{||}^2 + \left(\frac{m\pi}{L_c}\right)^2}, \tag{22}$$

where $c$ and $n$ denote the speed of light and the refractive index of the cavity material, respectively. Panel (a) of Fig.1 shows a cartoon of the cavity. Panel (b) demonstrates the excitation of a cavity mode by an external field when the frequency and incident angle match. Panels (c) and (d) show, as functions of $k_{||} = \left|\mathbf{k}_{||}\right|$, the cavity mode dispersion, Eq. (22) (dashed black line), the molecular exciton frequency (dashed-dotted beige) and the upper (blue) and lower (red) polaritons that result from their hybridization. In panel (c) the molecular frequency matches the cavity mode at the bottom of the band, $\omega_0 = \omega_c\left(k_{||} = 0\right)$, while in panel (d) the matching is at positive $k_{||}$, where $\omega_0 > \omega_c\left(k_{||} = 0\right)$. At these matching points (denoted by magenta dots in the figures) the spacing between the upper and lower polaritons is minimal and equal to $\Omega_N$.

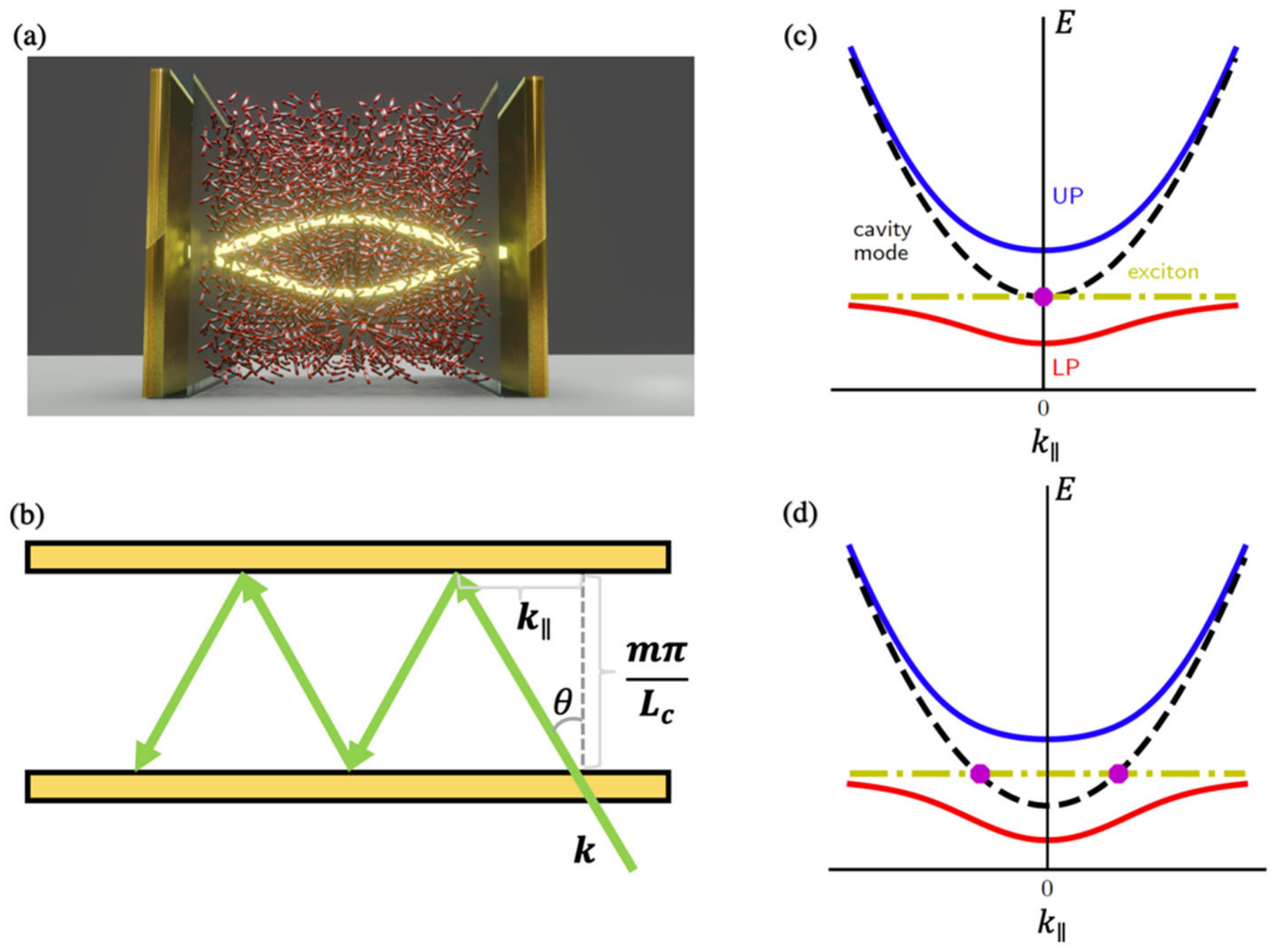

Figure 1. Polaritons in a Fabry-Pérot (FP) cavity. See text for details.

Several other points should be noted: First, as seen in Fig 1b the choice of $k_{||}$ can be done by the incident probe beam. For a given $k_{||}$, say $k_{||} = 0$ (normal incidence), the mode frequency can be tuned by modulating the spacing of the mirrors. Second, when considering the polaritons at a given $\omega_0$ and for varying $\omega_c$, maximum hybridization is achieved at resonance, $\omega_c = \omega_0$. As these frequencies depart, the modes become uncoupled, with the higher (lower) polariton branch becoming more photon-like (molecule-like) at $\omega_c > \omega_0$. The opposite holds when $\omega_c < \omega_0$. Finally, due to the translational symmetry along the mirror plane, polaritons with different in-plane wave vectors $k_{||}$ effectively do not interact with each other. (44) Therefore, in theoretical modeling of strong coupling in FP cavities one usually considers only a single cavity mode coupled to molecule, as modeled by the TC Hamiltonian. This simple consideration can sometimes breakdown. For example, if the system is continuously pumped and polaritons are excited, inelastic transitions between polaritons, $k_{||} \rightarrow k_{||}'$, may occur since polaritons have a molecular component and molecules do interact with each other. This may lead to intriguing phenomena like polariton condensate and polariton laser. (45) (46)

In FP cavities, light-matter coupling strength is determined by the mirror spacing as implied by Eq. (14), which puts a constraint on the maximum wavelength and therefore minimum cavity mode frequency. Alternatively, plasmonic cavities and related structures make it possible to achieve resonance conditions and strong coupling at sub-wavelength geometries. (2; 38; 43; 47) This has brought together research in strong light-matter coupling phenomena and in surface-enhanced spectroscopies, particularly surface-enhanced Raman scattering (SERS), where it was observed that such phenomena are often dominated by "hot-spots",(48; 49) essentially locations supporting strong light found at narrow gaps between metal nanostructures. Such plasmonic cavities (see Fig. 2), where plasmons replace cavity photons, were recently used in strong coupling studies down to a single molecule or a single nanodot, (50-53) while observations of Rabi splitting in FP cavities usually require the collective response of $10^6 - 10^{10}$ molecules. An important difference between FP and plasmonic cavities should be however kept in mind. In FP cavities, with typical dimensions in the micron range, the interaction of an individual molecule with the radiation field is relatively weak, and strong coupling reflects the collective response of many molecules. Other electrostatic interactions are not expected to change. In plasmonic coupling, the proximity of a metal interface may modify intermolecular and intramolecular electrostatic interactions and may affect molecular properties in a very different way.

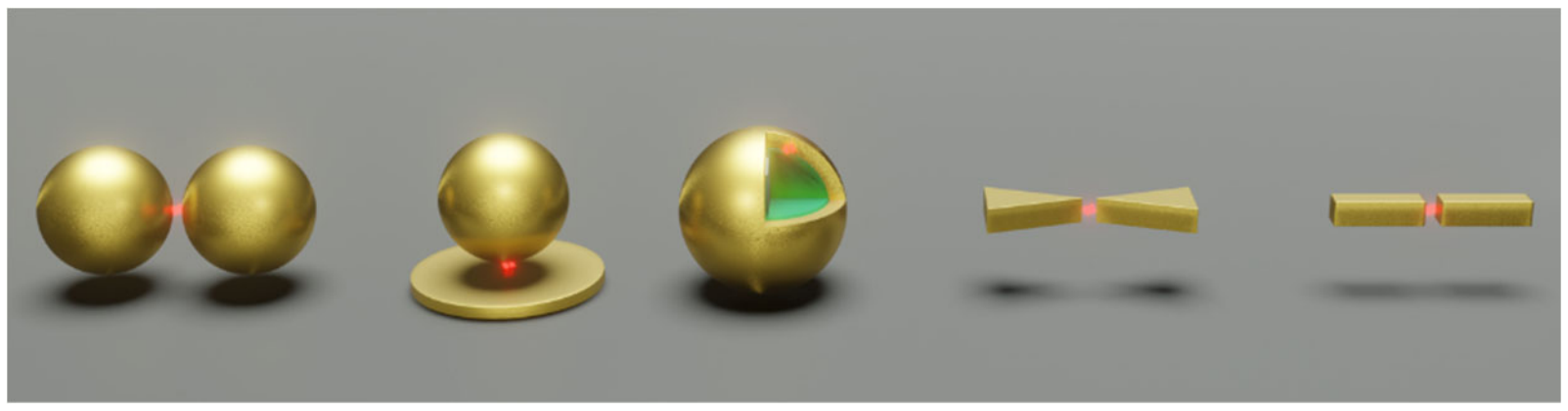

Figure 2. Different geometries of plasmonic "cavities": (from left to right) spheres, plane sphere, coated particles, bowties, and gaps between rods. The red emitter denotes the molecular cluster which forms strong coupling.

The hallmark of strong coupling behavior is the observation of Rabi splitting in absorption, emission, scattering, or reflection-transmission of the light used to probe the coupled light-matter system. Early observations of exciton-polaritons in semiconductor microcavities(54) (55) (56) where followed by similar observations using molecular dye aggregates (organic semiconductrs) in FP cavities(57) (58-60) as well as plasmonic cavities(61-63) (64) (65). Time resolved Rabi oscillations were observed, both in FP-like (66; 67) and plasmonic(68) configurations. For a recent review on exciton-plasmon strong coupling see(69). While the observed splitting strongly depends on the system and structures studies, values in the range of 0.5 eV were reported, see, e.g. (59) (60; 63) (64) and even values near and above 1eV were observed,(70) (71) approaching the limit of ultra-strong coupling. This has made it possible to observe this signature of strong coupling in plasmonic cavities despite the large dissipation rates associated with the metal interfaces. We note in passing that while the observation of Rabi splitting has been the obvious indicator of strong coupling in all the above works, the use of ellipsometric amplitude and phase spectra has been recently suggested as another signature.(72)

The studies cited above are examples of strong coupling to molecular electronic transitions. More recently, strong coupling to molecular nuclear motions has been demonstrated and intensively studied,(13; 14) (15) (16) (17) (18) (19-26) (73) with reported Rabi splittings in the range of 0.001-0.01 eV. Unlike exciton-polaritons, different molecular vibrational modes can be involved, leading to more complicated multi-mode Rabi splitting structures.(16) (74) (75). Most studies of VSC were done in FP-type cavities, however strong and ultrastrong coupling were recently observed also in plasmonic devices(76) (77) (78) (79).

For our later discussions, it is important to consider the number of molecules that participate in polariton formation. As mentioned above, in plasmonic cavities the coupling strength, sometimes expressed in terms of Eq. (14) with $\mathcal{V}$ representing an effective mode volume, is strong enough to observe the Rabi splitting indicating strong coupling with a single quantum emitter. In FP configurations the number of molecules $N$ involved in the polaritonic response is far larger. Using $\lambda^3$ as an estimate of the cavity volume, where $\lambda$ denotes the wavelength of the cavity photon that resonates with the molecular transition, and taking the intermolecular distance to be 1 nm, gives $N = 10^7$ and $N = 10^{10}$ for electronic and vibrational strong coupling, respectively, as order of magnitude estimates.

Observations of strong coupling in the optical response of molecules in confined optical environments naturally raise the possibility that other molecular processes such as charge and energy transport and chemical reaction rates may be affected. Several observations and related theoretical treatments have demonstrated and discussed the effect of coupling to cavity modes on

energy (80; 81) (82) (83) (84) (27) (85-89), charge(90) (91) (92-94) (95) and heat(96) transport in the last decade.[a] Conceptually, such effects are similar to earlier observations in which charge transfer along molecular bridges is affected by coupling to vibrational motions on the bridge(99-104). Particularly akin to the cavity phenomena are observations and discussions of the effect of vibrational motions on the coherence and range of exciton propagation in molecular systems.(105-112). The radiation field is another such bosonic environment. The continuum of radiation-field modes that are weakly coupled to matter in an open environment mainly provide a dissipative channel, however, the modes supported by an optically confined environment can couple to molecular processes in specific ways. Indeed, enhancement of exciton transport by coupling to surface plasmons has been extensively discussed, see, e.g. Refs. (113) (114) (115-117). Finally, as mentioned above, the Rabi splitting for exciton-polaritons can exceed $k_B T$ at the room temperature by more than an order of magnitude. Such a large Rabi splitting allows the realization of nonequilibrium Bose–Einstein polariton condensate and polaritonic laser at room temperature.(45) (46)

Some other interesting strong molecule-light coupling effects in optical cavities, such as an effect of the cavity vacuum field fluctuations on a metal work function(118) and on the non-linear optical response of cyanin dyes(119) (with electronic strong coupling), and some crystallization processes (under vibrational strong coupling)(120; 121). Electrochemical tuning of VSC was demonstrated in(122) and non-linear response of vibrational polaritons was explored in Refs. (123) (124). Arguably the most intriguing and potentially useful are effects pertaining to chemical reactivity which is discussed in more detail in the next Section.

## 3. Strong coupling effects on chemical dynamics

Broadly, chemical reaction dynamics is an expression of the interplay between nuclear and electron dynamics, where protons are sometimes treated at the same level as electrons. Within the Born-Oppenheimer picture, nuclear dynamics involves motions and barrier crossing on adiabatic potential surfaces, as well as non-adiabatic crossings between such surfaces, in the presence of energy exchange with the thermal environment. Electron (and sometimes proton) dynamics appear

---

[a] Note that cavity effect on heat transport is akin the many observations and discussion of radiative heat transfer in the near field limit.(97)(98)

as tunneling processes underlying the non-adiabatic transitions. Each of these fundamental constituents: the potential surfaces, energy exchange and relaxation, barrier crossing and non-adiabatic transitions including the quantum tunneling processes may be affected by strong coupling to optical cavity modes.

When considering possible cavity effects on chemical reaction dynamics we should distinguish between several factors. First, we must consider the difference between dark processes, where the cavity effect stems from its vacuum-field properties, and photochemical processes induced by external irradiation. Second, we should address the differences between FP and plasmonic cavities. In weak coupling situations, the advantage of the plasmonic cavity environment lies in the disconnect between mode energy and wavelength, making it possible to sustain modes that resonate with molecular electronic and even vibrational transitions in extreme subwavelength configurations and to study system sizes down to a single molecule. Under strong coupling, the proximity to metallic boundaries can alter chemical properties by affecting electrostatic interactions – an important effect that is not obviously related to polariton formation. There has been extensive work aimed at generalizing ab-initio quantum chemistry to the presence of strong radiation-field molecule coupling(125) (126) (127) (128) (129) (130) (86; 131) (132) (133; 134) (133) (135) (136; 137). It should be kept in mind, however, that processes in plasmonic environments may be dominated not by modifying the molecular electronic structure but by coupling to the dissipative metal boundaries thus yielding a dissipative channel that competes with an observed photochemical process,(138)[b] or directly affecting the metal boundaries giving rise, for example, to plasmon-induced hot electrons formation(139; 140) and affecting molecular electronic processes in this way. Multiscale QM/MM molecular dynamic simulations, (141) (142) (143) in which forces are calculated on the fly from a combination of electronic-structure calculations and mechanical forcefields and a surface hopping algorithm is used for transition between electronic states, are promising tools for describing such complex situations.

[b] Note that even in FP cavities, dissipation of the cavity photon constitutes a decay channel for molecular excitations via the polariton, which may reduce the yield of a competing photochemical process.

Next, relative timescales should be considered carefully. Our understanding of molecular physics largely stems from considerations based on the timescale separation between nuclear and electronic motions, leading to the Born-Oppenheimer approximation and the concept of nuclear potential surfaces, and discussions of non-adiabatic processes in terms of deviations from this underlying picture. Depending on physical parameters, the characteristic timescales associated with plasmon or cavity modes that couple to the molecular system may be similar to the electronic timescale, putting them in the group of fast electronic motions, or as slow as vibrational motion, whereupon they play the role of an additional nuclear coordinate. In either case, the result is a modified potential energy surface for the atomic motions in the molecule, but the implications may strongly differ as discussed below. Obviously, intermediate situations in which the inverse cavity mode frequency lies between the electronic and vibrational timescales are possible, however, such situations were rarely considered as the focus was on systems where the cavity mode is in resonance with characteristic electronic or nuclear molecular transitions. Finally, an intriguing issue is the distinction between single-molecule and collective effects. In the former, the cavity induced change reflects the effect of coupling to cavity mode(s) on the dynamical properties of individual molecules. In contrast, the observed collective optical response discussed in Section 1 has led to discussions of the possibility of collective chemical behavior. This may be a critical issue in FP-like systems, where the optical ramifications of strong coupling stem from the number of molecules that "see" a single optical mode, while the cavity-induced force experienced by a single molecule is much smaller than intramolecular or intermolecular interaction.

In the last decade, there have been several reports of experimental observations of cavity-induced chemical effects under strong coupling conditions. Only a few observations of chemical effects associated with strong coupling to molecular electronic transitions were reported so far, (140; 144-146) which is somewhat surprising given the sometimes dramatic theoretical predictions made regarding both individual molecular behaviors(147-149) (128) (133; 136) and possible molecular collective effects. (150) (151) (152) Note that the observation of Ref. (146) was made in a plasmonic cavity where, as noted above, chemical effects are not necessarily related to polariton formation.

While electronic strong coupling can potentially affect photochemical processes by modifying the potential surfaces on electronically excited states, VSC effects - strong coupling between molecular vibrational motions and cavity modes, has recently attracted attention following reports of "VSC catalysis", i.e., cavity effects on chemical reactions in the electronic ground state without external optical pumping. (153; 154) (155) (156) (157) (158) (159) (160) (see however (161)); see also Ref.(162) for a recent review. For example, studies(153) (157) of the desilylation of 1-phenyl-2-trimethylsilylacetylene (PTA; Fig 3a) show marked differences between the kinetics inside an FP cavity and in bulk liquid, with the former becomes slower in the cavity (Fig. 3b). Furthermore, the cavity kinetics is closely correlated with the cavity mode frequency (the deviation from bulk behavior is largest when a cavity mode at normal incidence ( $w_c\left(k_{\|}=0\right)$ ) is in resonance with the Si-C stretch vibration) and with the Rabi splitting when the latter is changed by changing the PTA concentration. Importantly, since the Rabi splitting is a collective effect, such correlation points to a collective origin of the cavity effect on this kinetics. An analysis of the temperature dependence of this correlation yields an intriguing correlation between the reaction free energy, enthalpy and entropy and the Rabi splitting(157). Another remarkable work from the same group (154) has shown that VSC can be used to control reaction pathways, enhancing some and suppressing others, by tuning cavity modes to different molecular vibrational modes. Also noteworthy is the observation(155) that if the reactant concentration is too small to support strong coupling, it is possible to catalyze the reaction (solvolysis of para-nitrophenyl acetate) by strong coupling to a vibration of a solvent which is in resonance with the reactant vibration.

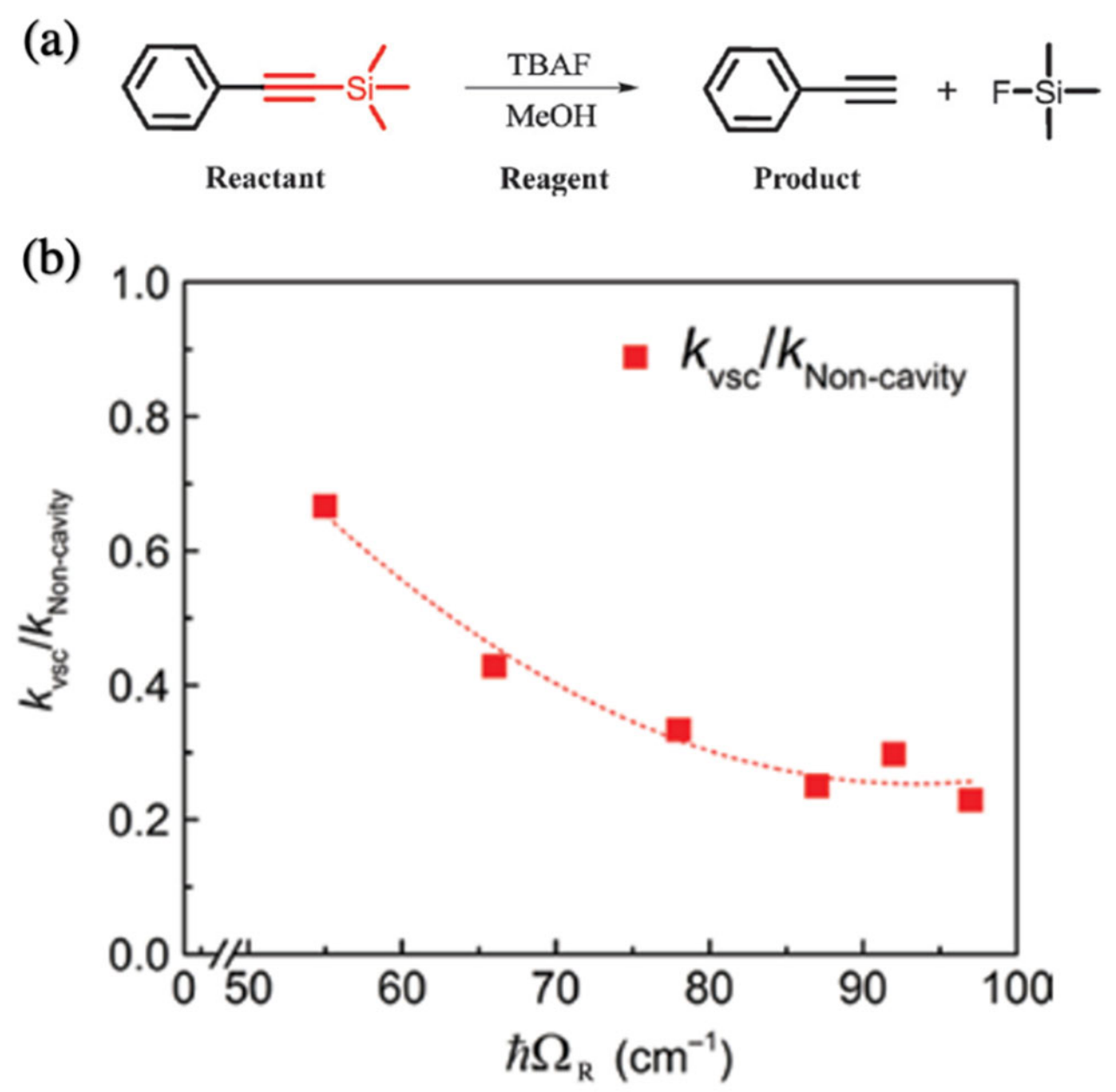

Fig 3. From Refs. (153) (157) The rate of PTA desilylation (a) is markedly different in and out of the cavity (b)

The fact that these observations were made in thermal systems in the dark does not rule out the possibility, raised in some theoretical papers, that these processes are dominated by underlying dynamical steps such as energy accumulation or transfer, solvation dynamics, barrier crossing, etc. Several recent works, using pump-probe(163) or 2-dimensional infrared (2d-IR)(124) (24) (26) spectroscopies, indeed indicate that VSC can affect molecular relaxation processes. It was found that the upper polariton relaxes on a ps timescale to vibrational dark modes.(163) (24) (26) However, the relaxation of the lower polariton is more complex and appears to be faster (< 5 ps) than the rate at which dark states are populated (up to 30 ps). (124), suggesting the involvement of some intermediate state in this relaxation(124). It should be noted that there are no indications that such non-equilibrium phenomena, that follow optical excitation of the system are implicated in the observations of "VSC catalysis" in Refs. (153; 154) (155) (156) (157) (158) (159) that occurs in thermal conditions.

### 3a. Theoretical considerations: Electronic strong coupling.

Photochemical processes are usually initiated by photon absorption by the ground state molecule. The subsequent nuclear dynamics on the potential surface of the excited electronic states may lead to non-adiabatic transitions to other electronic states (as is usually the case for photoinduced electron transfer) and eventually to the photochemical product. Theoretical considerations of ways by which this scenario may be affected by coupling to cavity mode(s) have taken place along several observations:

(a) Modification of molecular potential surfaces.(148; 164) (141; 147) (128; 129; 131; 149) (133; 136) A cavity mode in resonance with an electronic transition is a fast mode that can be used as part of the electronic sub-system when evaluating ground and excited-state potential surfaces. As a function of the nuclear configuration, this change is local because the molecular electronic energy spacing ($\omega_0$ in Eq.(2) or (20)) strongly depends on this configuration. The resulting polaritonic potential surface can be very different from the corresponding molecular surfaces in free space, altering the course of the nuclear motions that lead to a chemical reaction.

(b) Electron vibration decoupling in the large *N* limit. It should be noted that the above is a single molecule picture that requires strong coupling on the level of one or a few molecules, a possible scenario for a plasmonic cavity. If a Rabi splitting of, say, 0.5 eV reflects coupling of $10^7$ molecules in an FP cavity, the coupling per molecule is of the order of $10^{-3}$ ev and its effect on a single molecule behavior should be carefully scrutinized. At issue is the nature of the Born-Oppenheimer potential surfaces in polaritons that involve a large number of molecules. Several workers(147; 150; 164-167) have addressed this problem using an extension of the Tavis-Cummings model of Eq. (16) based on the Holstein polaron model.(168; 169) In the extended mode, each two-level species is replaced by a “molecule” that comprises two electronic states and one harmonic oscillator, representing a molecular vibration, with the standard polaronic interaction (namely, a horizontal shift of the harmonic surface) between them. This is described by the Holstein-Tavis-Cummings Hamiltonian

$$\hat{H}_{HTC} = \hbar\omega_c \hat{a}^\dagger \hat{a} + \hbar\sum_{j=1}^{N}\left[\omega_0 \hat{\sigma}_j^+ \hat{\sigma}_j^- + g_0\left(\hat{a}^\dagger \hat{\sigma}_j^- + \hat{a}\hat{\sigma}_j^+\right) + \omega_v\left(\hat{b}_j^\dagger \hat{b}_j - \lambda_0 \hat{\sigma}_j^+ \hat{\sigma}_j^-\left(\hat{b}_j^\dagger + \hat{b}_j\right)\right)\right] + \sum_{j=1}^{N}\sum_{i(\neq j)=1}^{N} J_{ij}\hat{\sigma}_i^+ \hat{\sigma}_j^- \qquad (23)$$

The terms linear in $\omega_v$ in Eq. (23) represent the molecular vibration and its coupling to the electronic subsystem, respectively, while the last term (usually representing dipole-dipole coupling between molecules) that describe exciton hoping between molecular sites has been disregarded in the TC Hamiltonian (21).

Consider first the Holstein Hamiltonian, Eq. (23), without the terms associated with the cavity, applied to exciton or electron hopping on a 1-dimensional chain with nearest neighbor hopping: $\hat{H}_H = \hbar\sum_{j=1}^{N}\left[\omega_0\hat{\sigma}_j^+\hat{\sigma}_j^- + \omega_v\left(\hat{b}_j^\dagger\hat{b}_j - \lambda_0\hat{\sigma}_j^+\hat{\sigma}_j^-\left(\hat{b}_j^\dagger + \hat{b}_j\right)\right)\right] + J\sum_{i=1}^{N}\hat{\sigma}_{j+1}^+\hat{\sigma}_j^-$. This Hamiltonian is often used as a model for charge or energy transport on a 1-dimensional tight-binding chain with nearest neighbor couplings. The effect of the nuclear motions represented by the $\hat{b},\hat{b}^\dagger$ operators and this transport process depends on the relative magnitudes of the polaron coupling $\lambda_0$ (or the reorganization energy, $E_R = \hbar\omega\lambda_0^{\,2}$) and the intersite coupling *J*. When the exciton hopping elements *J* are small, transition rates between molecular sites are strongly affected by overlap between nuclear wavefunction, which scales like $\exp\left(-\lambda_0^{\,2}\right)$ when the hopping is between ground states of the vibrations in the expressions for transition rates. For large *J*, these overlap integrals become less important and scale like $\exp\left(-\lambda_0^{\,2}/N\right)$,(170) (166) where *N* is the coherence length (expressed in number of sites) of the exciton. In molecular J-aggregates *N* is much smaller than the actual number of molecular chromophores. However Spano(166) has pointed out that strong exciton-cavity photon coupling effectively eliminates the effect of disorder, making *N* much larger and the vibrational overlap essentially unity, leading to an effective decoupling between electronic and vibrational dynamics in the lower polariton band. Qualitatively this can be explained by noting that the polariton Born-Oppenheimer surface, namely the electronic energy associated with the bright state $|B\rangle$ of Eq. (7) is given by $N^{-1}\left[\sum_j E_x\left(R_j\right) + \sum_{k\neq j} E_g\left(R_k\right)\right]$, which implies that each nuclear coordinate $R$ experiences a potential surface $N^{-1}E_x\left(R\right) + \left(\left(N-1\right)/N\right)E_g\left(R\right)$, namely making the polaritonic and ground state potentials essentially the same for large N. Herrera and Spano(150) have taken this idea further, suggesting that the rate of a photo-induced electron transfer can be dramatically different from that associated with a single excited molecule, when this excited state is part of a polariton formed by many-molecules interacting with a cavity photon. This difference reflects again the observation that in the polaritonic state the vibronic coupling is expressed far more weakly than in the single molecule. This issue was further studied in Refs. (171; 172) (164; 165) (167) concluding that vibronic decoupling holds only provided that the Rabi splitting $\Omega_N$ is large relative to the reorganization energy $E_R$. Even in this restrictive form, such decoupling can have far-reaching consequences as has been pointed out by several recent theoretical works. (150) (147; 151) (152; 172; 173)

(c) Effect of disorder. The consequences of collective coupling, as described above, may depend, sometimes strongly, on disorder and dephasing. In discussing disorder we usually distinguish between the extreme limits of static disorder on one hand, and fast dynamic disorder on the other. For example, static disorder gives rise to inhomogeneous broadening while the fast limit of dynamic disorder translates in the Markovian limit of kinetic equations to relaxation and dephasing rates that are expressed also in the homogeneous broadening of spectral features. Here the "fast" and "slow" (or "static") qualities are attributed relative to the observed timescale. Intermediate situations obviously exist and are addressed by using non-Markovian descriptions of the system dynamics or by increasing the number of relevant system variables.

As mentioned above, another manifestation of disorder and dephasing effects is in limiting the coherence length in the response of molecular systems, roughly defined as a characteristic distance within which molecules behave collectively. In the optical processes discussed in the review, the interplay between disorder and collective response can be looked at the consequences of competing phenomena – optical interactions that affect the molecules collectively, and disorder that disrupts the collective response. This issue has been addressed in studies of the way static disorder affects collective radiative emission, (174) (175) (176-179) leading to the conclusion that the behavior of the superradiance effect (both the incubation time for collective decay on the disorder and the nature of the emission) depends on the amplitude of disorder compared with the strength of the light-molecule coupling (which, in turn, can be controlled by a cavity environment). Another manifestation of such behavior is seen by the observation, both experimental(14) (180) and computational(181), that an inhomogeneously-broadened molecular absorption line splits, under strong coupling with a resonating cavity mode, into upper and lower polaritons peaks that are "immune" to the inhomogeneous broadening of the corresponding molecular transition outside the cavity.

**3b. Theoretical considerations: Vibrational strong coupling.**

When the cavity supports modes that are close to resonance with molecular vibrations these modes become part of the forcefield that defines the nuclear potential surface, which in turn affects the nuclear motion of a single molecule and may potentially (since the nuclear motions of many molecules are coupled to the same cavity mode) result in the collective molecular behavior. These possible ramifications of VSC were explored in recent theoretical works along several lines:

(a) Transition state theory (TST, see, e.g. (182)) has been successful in describing thermal chemical rate processes in condensed phases. The TST rate is an approximation based on two assumptions: first, that thermal equilibrium is maintained in the reactant(s) subspace, and second, that once the transition state configuration is achieved, escape into the product subspace is irreversible. It can be shown(182) that these assumptions lead to an upper bound to the actual rate. Several works have applied this level of rate theory to a molecular system under VSC conditions, with the disappointing result that the cavity environment does not make a significant difference in the TST rate in the large $N$ (number of coupled molecules) limit(183) (184) (185; 186), that is, in the FP configuration.[c] A way to show this(184) within classical transition state theory examines the potential of mean force (PMF) along the reaction coordinate $x_r$

$$\mathrm{PMF}(x_r) = \frac{\int d\mathbf{x}d\mathbf{p}\,\delta(\mathbf{x}-x_r)\,e^{-\beta H}}{\int d\mathbf{x}d\mathbf{p}\,e^{-\beta H}} \tag{24}$$

where, $\mathbf{x}$ and $\mathbf{p}$ denote the positions and momenta of all particles (nuclei + cavity modes), $x_r$ denotes the reaction coordinate, and $H$ denotes the Hamiltonian. If we (i) assume classical nuclei and photons (ii) ignore the cavity modification of electrostatic interactions and the molecular potential surface (which is valid for FP cavities), the resulting PMF is the same inside versus outside the cavity and no VSC catalytic effect would be expected.(184) It should be noted that this conclusion is based on a representation of the molecule-radiation field interaction that includes self-dipole interactions (see also Ref. (187)) – a standard form in any exact theory, whose application for truncated molecular models has been recently questioned.(37)

(b) When the molecule-cavity mode coupling is large enough, as would be the case in plasmonic cavities constructed to support modes in the IR domain, interesting effects are predicted for classical thermal reactions(183) (188) (95) (158) (189) (190) as well as electron transfer rates.(86) (191) (192) Such predictions may be relevant for molecules in plasmonic cavities (if constructed to support modes in the IR domain), but appear to be inconsequential for the effects observed in FP environments.

---

[c] Given the experimentally known magnitude of the Rabi splitting, which is proportional to $\sqrt{N}$, "large $N$ limit" should be understood as "large $N$ limit for a given Rabi splitting, namely interaction per molecule that scales as $g \sim N^{-1/2}$.

(c) The realization that observed VSC effects on chemical rates can not be explained by equilibrium rate theory has prompted efforts to examine special circumstances, e.g. Galego et al(183) have found that a cavity catalytic effect emerges only when the molecules have permanent dipoles that have a specific orientation relative to the cavity mirror. Also, possible interesting effects were found in electron transfer reactions where the transfer is accompanied by large changes of spectral properties so only one side of the reaction is strongly coupled to the cavity mode (193) (see also (194)). Such specific models and others(195) cannot easily be adapted to current experimental observations.

(d) The failure of equilibrium rate theory (TST) observation has also prompted examinations of possible non-equilibrium effects. As discussed above, cavity effects on transient molecular response under VSC are indeed observed by pump-probe and 2d-IR spectroscopies(163) (124) (24) (26) and theoretical treatments of such observations were advanced.(196) (197) (198) In particular, it has been pointed out that, because many molecules share coupling to the same cavity mode, intermolecular vibrational energy transfer can be enhanced in the cavity environment. (21) (199) (200) (201) (202) This, in turn, may have implications on chemical rates: increasing the rate of relaxation of nuclear motion may increase or decrease the rate of a chemical reaction in different limits of coupling of this motion to the thermal environment, however, these limits are usually observed in the gas phase (where the chemical rate increases with nuclear ) or in highly viscous solutions (decreased rate). Furthermore, it should be kept in mind that any possible effect of coupling to a cavity mode on nuclear relaxation comes on top of the relaxation induced by intermolecular interactions. In FP cavities the latter are often dominant.

(e) One difficulty encountered when trying to bridge the gap between experimental observations of what appears to be collective behavior in chemical reaction rates in VSC situations and current theoretical understanding is the fact that unlike the molecular response to an optical probe, chemical events are local. For example, we can picture $N$ molecules responding together to light whose coherence length encompasses these molecules, but it is hard to picture $N$ molecules crossing a barrier together: for nothing else, the needed activation energy will make such a process impossible. One possible exception involves the interaction between a polariton and an impurity species embedded in the same environment. Consequences of such interaction were suggested as a way to interpret the experimental observations of Ref. (155) and were observed computationally,(203) (204). For example, classical molecular dynamic simulations(204) show that such interaction can provide an efficient way to accumulate energy in a reaction coordinate by energy transfer from an excited polariton mode. (205) (206) (159) It remains to be seen whether an analog of such a process can be active in a thermal system.

(f) Since VSC involves interaction between nuclear motions in the ground molecular electronic and harmonic cavity modes, its main features can be accounted for by classical MD simulations. Li et al have recently developed a cavity molecular dynamics (CavMD) scheme, (207) (208) (201) (204) which propagates the coupled dynamics between a few cavity modes (treated classically) molecular systems in the electronic ground-state surface with classical force fields. CavMD has been shown to recover (i) polariton relaxation (or dephasing) to vibrational dark modes on a timescale of ps or sub-ps, (208) and (ii) polariton-enhanced molecular nonlinear absorption when the LP is strongly excited, (208) a signature of which is a delayed population gain in the first excited state of vibrational dark modes after the LP excitation.(124) These simulations also provide numerical insights on understanding VSC catalysis.(207) (201) For example, for liquid water under VSC, CavMD shows that individual molecular properties are either entirely unchanged or only weakly modified under thermal conditions,(207) in agreement with the results from transition state theory presented above, and indicating again that "VSC catalysis" may have a non-equilibrium or quantum origin. Such simulations also provide insight into the effect of system size: changing the number of interacting molecules $N$ while keeping the Rabi splitting constant (which amount to changing the cavity volume at constant molecular density) make it possible to distinguish between inherently collective effects (such as the Rabi splitting) and those that reflect individual molecular behaviors (such as barrier crossing). The latter usually scale like $1/N$ in such a limiting process. Potentially important is the observation that energy transfer processes can show collective behavior. A recent nonequilibrium CavMD simulation(201) demonstrates that for a small fraction of vibrationally excited guest molecules immersed in a liquid thermal bath, vibrational energy transfer from transient polaritonic excitations in the gest to the host molecules scales like $N^{-x}$ with $x<1$, in addition to showing a resonance dependence on the cavity mode detuning. Furthermore, because of molecular anharmonicity, this energy transfer process results in transient distributions with an excess population in the high energy tail, indicating a possible effect on chemical rates (that are dominated by the same high energy tail). Such observations, as well as higher-level quantum calculations,(137) come closer to the experimental results of Ref. (155) but still falls short of rationalizing it: The observed $N^{-x}$ with $x \sim 0.7$ in the particular system studied in (201) would predict a negligible cavity effect per molecule for $N > 10^4$ and currently it is unclear that whether the more prominent cavity effect in the high energy tail would persist in larger cavities.

(h) In conjunction with strong coupling phenomena, Raman scattering can be considered from two perspectives: First, is the effect of VSC experienced by the Raman mode. (209) (210) (211) Second, one can study resonance Raman scattering in situations where the intermediate resonance electronic state is strongly coupled to a molecular cavity mode. (212) (213) (214) (215) As in observations of cavity effects on chemical rates, an apparent discrepancy between observations of large cavity enhancement of the Raman signal(209) and theory(210) that fails to see this enhancement (consistent with another observation(211)) needs to be resolved. Underlying both types of observations is the question how coherent and collective is spontaneous Raman in such systems(215) as compared to Rayleigh scattering. (216)

## 4. Energy conversion

The recognition that an optical cavity environment is a tool for controlling chemical and photochemical processes naturally raises the possibility of applications in energy conversion. We note that the active field of cavity optomechanics(217) provides another perspective on this issue. Also, the use of plasmonic particles to enhance the performance of photovoltaic cells has been considered: such particles can be used to amplify the local electromagnetic field and also as sources of plasmon-induced hot electrons. (218; 219) Strong coupling of molecules to cavity polaritons can affect energy transfer via polariton formation in more specific ways,(206) and has been suggested as a way to control singlet fission(220) and triplet energy harvesting.(221) Indeed, polariton formation has been implicated in recent observation of triplet pair annihilation, (222; 223) intersystem crossing(224) and internal conversion processes.(173)

While still on the purely theoretical level, it is worthwhile to mention in this context a recent effort to analyze the implication of collective quantum transport on the efficiency and rate of energy accumulation and work extraction, so called quantum batteries(225) with potential applications to the performance of engines where a molecular system coupled to a cavity photon operates as the working fluid. Alicki and Fannes(226) have posed the question what is the maximum work that can be extracted from a system that starts at some given state, and have made the interesting observation that this work increases, then saturates with the number $N$ of copies of the "engines" used. While the original assertion that this behavior originates from quantum entanglement between the copies was repudiated,(227; 228) this observation has led to further examination of the possibility to enhance thermodynamic performance by collective response. (229) (230) (225) In particular, the Dicke superradiance concept has been generalized, using a Tavis-Cumming model of a quantum battery, to show that it is possible to get enhanced power - accelerated charging and discharging of such systems.(230) (231) A cavity environment can provide a tool to control energy storage and recovery in such devices.(232) (233; 234) Such constructs were recently explored as the working components in full heat engine models. (235) (236) It remains to be seen whether these ideas will eventually go from concept to practice.

## 5. Technical issues: Gauge and Hamiltonian for modeling strong coupling

When modeling strong coupling, an issue concerning the appropriate form of the light-matter Hamiltonian often comes up. This question arises because when dealing with light-matter interactions, there is a freedom to add an arbitrary gauge function on top of the light-matter Lagrangian and write down the corresponding Hamiltonian under this gauge. (29) Popular gauges, in nonrelativistic QED, include the Coulomb gauge (where the light-matter coupling is $\mathbf{p}\cdot\mathbf{A}$) and the Power-Zienau-Woolley gauge (or the dipole gauge, where the light-matter coupling is $\boldsymbol{\mu}\cdot\mathbf{E}$ ).(29)

In principle, since a gauge transform is a unitary transformation and should not change the value of any physical observables, any gauge can be chosen, so one usually uses a gate that is convenient for solving the problem at hand. Hence, in principle, either $\mathbf{p}\cdot\mathbf{A}$ (in the Coulomb gauge) or $\boldsymbol{\mu}\cdot\mathbf{E}$ (in the dipole gauge) can be used to express the light-matter coupling in study of strong coupling phenomena. This statement, however, is sometimes problematic when other approximations are made. During the past decade, this gauge ambiguity issue has come under intense discussion. (30) (31) (32) (33). One main conclusion is that one cannot simply ignore the $\mathbf{A}^2$ term (in the Coulomb gauge) or dipole self-energy term (or the self-dipole term in the dipole gauge). If the $\mathbf{A}^2$ or the self-dipole term is ignored, one can get surprising conclusions, including a gauge-dependent solution and energy divergence,(31) or an instability of electronic ground state,(34) especially when the light-matter coupling is large and in the ultrastrong coupling limit. Within certain truncated bases, however, it is possible that this term might disappear in a particular gauge so that one may obtain promising results even without adding the $\mathbf{A}^2$ or the self-dipole term, e.g., the quantum Rabi model.(31) Another issue is that, even when the $\mathbf{A}^2$ or the self-dipole term is included in the Hamiltonian, since realistic calculations usually use a truncated basis, a good choice of gauge might reduce the computational error.(35) Additionally, the usually assumed rotating wave approximation (e.g., in the TC Hamiltonian) becomes problematic when the light-matter coupling becomes strong, and the inclusion of the counter rotating wave terms are important. (28) Hence, it is crucial to use an appropriate Hamiltonian to deal with strong coupling.

The discussions in this review were mostly presented using two-level system(s) instead of realistic molecules. From a chemist's point of view, the literature of gauge ambiguities for realistic molecules is still very limited(34) (36) and there is still some debate on the appropriate Hamiltonian, e.g., for plasmonic strong coupling.(37) It appears that working under the dipole-gauge Hamiltonian (with the self-dipole term included) seems to be a convenient (and perhaps safe) way to deal with light-matter coupling in Fabry–Pérot environments. (207) (128) (184) (34) Future work is necessary to assess the issue of gauge ambiguities for realistic molecules.

## 6. Challenges and opportunities

Current theoretical understanding of strong coupling phenomena, both in the electronic and vibrational domains of molecular physics is still limited. Theoretical analysis of electronic strong coupling phenomena appears to be well developed but the lack of experimental verification of very profound predictions should perhaps be of concern. In the VSC domain we face a gap between experimental observations of "VSC catalysis" and theoretical understanding that will hopefully close in the near future. For the advanced experimental techniques in this domain - pump-probe and 2D-IR spectroscopies of vibrational polaritons, it is challenging to give a quantitative theoretical description when the experimental setup becomes complicated. (123) (83) (237)

Focusing on VSC catalysis in thermal reactions, although a few theoretical works seem to suggest that considering nonequilibrium dynamics could yield a resonant effect on individual molecules, (201) (190) (137) experiments in FP cavities indicate that only the resonance with the cavity mode at $k_{\parallel} = 0$ (Fig. 1c, as opposed to Fig. 1d) is meaningful. This distinction cannot be rationalized by current theoretical treatments. At the same time, the observed collective effect in observed chemical rates appears to be at odds with the theoretical prediction of $1/N$-scaling of such phenomena. (201) (203) When IR photochemical reactions are considered, one possible route towards VCS enhancement of photochemical rates is the observation that chemical reactions in the electronic ground state can be induced by IR pumping of the reaction coordinate(238) (239) together with the fact, discussed above, that polaritons excited in the solvent can efficiently transfer energy to (possibly reactive) solute species. (204)

The Raman spectroscopy issue shares some similarities with VSC catalysis - both are usually considered as a statistical average of individual molecular events, leading to an expected negligible collective effect in a FP cavity, which conflicts with some experiments. As the Raman response is easier to calculate than a chemical rate, clarifying the Raman issue might shed light on understanding VSC catalysis. It is interesting to note that recent experiments outside a cavity appear to suggest that there could be collective quantum coherence during the spontaneous Raman scattering process in the liquid phase, (240) which might facilitate our understanding of the Raman spectroscopy issue under VSC.

Finally, the possible use of strong coupling phenomena in affecting and controlling molecular transport processes, offers a rich field for further explorations of using such systems as energy pumping, storage transfer and conversion devices. An optical cavity provides a flexible and controllable environment, potentially enhanced by collective molecular response. Further research in this direction is obviously warranted.

**Acknowledgements.** This work has been supported by the U.S. National Science Foundation under Grant No. CHE1953701( AN) and by the U.S. Department of Energy, Office of Science, Office of Basic Energy Sciences, under Award No. DE-SC0019397 (JES). It also used resources of the National Energy Research Scientific Computing Center (NERSC), a U.S. Department of Energy Office of Science User Facility operated under Contract No. DE-AC02-05CH11231.

1. Novotny L. 2010. Strong coupling, energy splitting, and level crossings: A classical perspective. *American Journal of Physics* 78:1199-202
2. Sukharev M, Nitzan A. 2017. Optics of exciton-plasmon nanomaterials. *J. Phys. Cond. Mat.* 29:443003
3. Bricks JL, Slominskii YL, Panas ID, Demchenko AP. 2017. Fluorescent J-aggregates of cyanine dyes: basic research and applications review. *Methods and Applications in Fluorescence* 6:012001
4. Dicke RH. 1954. Coherence in Spontaneous Radiation Processes. *Phys. Rev.* 93:99-110

5. Gross M, Haroche S. 1982. Superradiance: An essay on the theory of collective spontaneous emission. *Physics Reports* 93:301-96
6. Lim S-H, Bjorklund TG, Spano FC, Bardeen CJ. 2004. Exciton Delocalization and Superradiance in Tetracene Thin Films and Nanoaggregates. *Physical Review Letters* 92:107402
7. Tavis M, Cummings FW. 1968. Exact Solution for an $N$-Molecule---Radiation-Field Hamiltonian. *Phys. Rev.* 170:379-84
8. Tavis M, Cummings FW. 1969. Approximate Solutions for an $N$-Molecule-Radiation-Field Hamiltonian. *Phys. Rev.* 188:692-5
9. Jaynes ET, Cummings FW. 1963. Comparison of quantum and semiclassical radiation theories with application to the beam maser. *Proceedings of the IEEE* 51:89-109
10. Shore BW, Knight PL. 1993. The Jaynes-Cummings Model. *J. Mod. Opt.* 40:1195-238
11. Flick J, Appel H, Ruggenthaler M, Rubio A. 2017. Cavity Born–Oppenheimer Approximation for Correlated Electron–Nuclear-Photon Systems. *Journal of Chemical Theory and Computation* 13:1616-25
12. Fábri C, Halász GJ, Cederbaum LS, Vibók Á. 2021. Born–Oppenheimer approximation in optical cavities: from success to breakdown. *Chem. Sci.* 12:1251-8
13. Shalabney A, George J, Hutchison J, Pupillo G, Genet C, Ebbesen TW. 2015. Coherent coupling of molecular resonators with a microcavity mode. *Nat. Commun.* 6:5981
14. Long JP, Simpkins BS. 2015. Coherent Coupling between a Molecular Vibration and Fabry-Perot Optical Cavity to Give Hybridized States in the Strong Coupling Limit. *ACS Photonics* 2:130-6
15. George J, Shalabney A, Hutchison JA, Genet C, Ebbesen TW. 2015. Liquid-Phase Vibrational Strong Coupling. *The Journal of Physical Chemistry Letters* 6:1027-31
16. George J, Chervy T, Shalabney A, Devaux E, Hiura H, et al. 2016. Multiple Rabi Splittings under Ultrastrong Vibrational Coupling. *Phys. Rev. Lett.* 117:153601
17. Simpkins BS, Fears KP, Dressick WJ, Spann BT, Dunkelberger AD, Owrutsky JC. 2015. Spanning Strong to Weak Normal Mode Coupling between Vibrational and Fabry-Perot Cavity Modes through Tuning of Vibrational Absorption Strength. *ACS Photonics* 2:1460-7
18. Casey SR, Sparks JR. 2016. Vibrational Strong Coupling of Organometallic Complexes. *The Journal of Physical Chemistry C* 120:28138-43
19. Vergauwe RMA, George J, Chervy T, Hutchison JA, Shalabney A, et al. 2016. Quantum Strong Coupling with Protein Vibrational Modes. *J. Phys. Chem. Lett.* 7:4159-64
20. Erwin JD, Smotzer M, Coe JV. 2019. Effect of Strongly Coupled Vibration–Cavity Polaritons on the Bulk Vibrational States within a Wavelength-Scale Cavity. *The Journal of Physical Chemistry B* 123:1302-6
21. Dunkelberger AD, Spann BT, Fears KP, Simpkins BS, Owrutsky JC. 2016. Modified relaxation dynamics and coherent energy exchange in coupled vibration-cavity polaritons. *Nat. Commun.* 7:13504
22. Ahn W, Vurgaftman I, Dunkelberger AD, Owrutsky JC, Simpkins BS. 2018. Vibrational Strong Coupling Controlled by Spatial Distribution of Molecules within the Optical Cavity. *ACS Photonics* 5:158-66
23. Dunkelberger AD, Davidson RB, Ahn W, Simpkins BS, Owrutsky JC. 2018. Ultrafast Transmission Modulation and Recovery via Vibrational Strong Coupling. *J. Phys. Chem. A* 122:965-71

24. Xiang B, Ribeiro RF, Dunkelberger AD, Wang J, Li Y, et al. 2018. Two-dimensional infrared spectroscopy of vibrational polaritons. *Proceedings of the National Academy of Sciences* 115:4845
25. Dunkelberger AD, Grafton AB, Vurgaftman I, Soykal OO, Reinecke TL, et al. 2019. Saturable Absorption in Solution-Phase and Cavity-Coupled Tungsten Hexacarbonyl. *ACS Photonics* 6:2719-25
26. Grafton AB, Dunkelberger AD, Simpkins BS, Triana JF, Hernández FJ, et al. 2021. Excited-state vibration-polariton transitions and dynamics in nitroprusside. *Nat. Commun.* 12:214
27. Feist J, Garcia-Vidal FJ. 2015. Extraordinary Exciton Conductance Induced by Strong Coupling. *Phys. Rev. Lett.* 114:196402
28. Frisk Kockum A, Miranowicz A, De Liberato S, Savasta S, Nori F. 2019. Ultrastrong coupling between light and matter. *Nature Reviews Physics* 1:19-40
29. Cohen-Tannoudji C, Dupont-Roc J, Grynberg G. 1997. *Photons and Atoms: Introduction to Quantum Electrodynamics*. New York: John Wiley & Sons
30. Vukics A, Grießer T, Domokos P. 2014. Elimination of the $A$-Square Problem from Cavity QED. *Physical Review Letters* 112:073601
31. Di Stefano O, Settineri A, Macrì V, Garziano L, Stassi R, et al. 2019. Resolution of gauge ambiguities in ultrastrong-coupling cavity quantum electrodynamics. *Nature Physics* 15:803-8
32. Stokes A, Nazir A. 2021. Implications of gauge-freedom for nonrelativistic quantum electrodynamics. *arXiv:2009.10662*
33. Settineri A, Di Stefano O, Zueco D, Hughes S, Savasta S, Nori F. 2021. Gauge freedom, quantum measurements, and time-dependent interactions in cavity QED. *Physical Review Research* 3:023079
34. Schäfer C, Ruggenthaler M, Rokaj V, Rubio A. 2020. Relevance of the Quadratic Diamagnetic and Self-Polarization Terms in Cavity Quantum Electrodynamics. *ACS Photonics* 7:975-90
35. Stokes A, Nazir A. 2019. Gauge ambiguities imply Jaynes-Cummings physics remains valid in ultrastrong coupling QED. *Nat. Commun.* 10:499
36. Taylor MAD, Mandal A, Zhou W, Huo P. 2020. Resolution of Gauge Ambiguities in Molecular Cavity Quantum Electrodynamics. *Physical Review Letters* 125:123602
37. Feist J, Fernández-Domínguez AI, García-Vidal FJ. 2021. Macroscopic QED for quantum nanophotonics: emitter-centered modes as a minimal basis for multiemitter problems. *Nanophotonics* 10:477-89
38. Törmä P, Barnes WL. 2015. Strong coupling between surface plasmon polaritons and emitters: a review. *Reports on Progress in Physics* 78:013901
39. Ruggenthaler M, Tancogne-Dejean N, Flick J, Appel H, Rubio A. 2018. From a quantum-electrodynamical light–matter description to novel spectroscopies. *Nat. Rev. Chem.* 2:0118
40. Herrera F, Owrutsky J. 2020. Molecular polaritons for controlling chemistry with quantum optics. *Journal of Chemical Physics* 152
41. Pascual JG. 2020. *Polaritonic Chemistry: Manipulating Molecular Structure Through Strong Light–Matter Coupling*. Springer
42. Kéna-Cohen S, Yuen-Zhou J. 2019. Polariton Chemistry: Action in the Dark. *ACS Central Science*

43. Climent C, Garcia-Vidal FJ, Feist J. 2021. CHAPTER 10 Cavity-modified Chemistry: Towards Vacuum-field Catalysis. In *Effects of Electric Fields on Structure and Reactivity: New Horizons in Chemistry*:343-93: The Royal Society of Chemistry. Number of 343-93 pp.
44. Agranovich VM, Litinskaia M, Lidzey DG. 2003. Cavity polaritons in microcavities containing disordered organic semiconductors. *Phys. Rev. B* 67:085311
45. Deng H, Haug H, Yamamoto Y. 2010. Exciton-polariton Bose-Einstein condensation. *Rev. Mod. Phys.* 82:1489-537
46. Keeling J, Kéna-Cohen S. 2020. Bose–Einstein Condensation of Exciton-Polaritons in Organic Microcavities. *Annual Review of Physical Chemistry* 71:435-59
47. Hugall JT, Singh A, van Hulst NF. 2018. Plasmonic Cavity Coupling. *ACS Photonics* 5:43-53
48. Campion A, Kambhampati P. 1998. Surface-enhanced Raman scattering. *Chemical Society Reviews* 27:241-50
49. Langer J, Jimenez de Aberasturi D, Aizpurua J, Alvarez-Puebla RA, Auguié B, et al. 2020. Present and Future of Surface-Enhanced Raman Scattering. *ACS Nano* 14:28-117
50. Santhosh K, Bitton O, Chuntonov L, Haran G. 2016. Vacuum Rabi splitting in a plasmonic cavity at the single quantum emitter limit. *Nat. Commun.* 7:ncomms11823
51. Chikkaraddy R, de Nijs B, Benz F, Barrow SJ, Scherman OA, et al. 2016. Single-molecule strong coupling at room temperature in plasmonic nanocavities. *Nature* 535:127-30
52. Groß H, Hamm JM, Tufarelli T, Hess O, Hecht B. 2018. Near-field strong coupling of single quantum dots. *Sci. Adv.* 4:eaar4906
53. Kongsuwan N, Demetriadou A, Chikkaraddy R, Benz F, Turek VA, et al. 2018. Suppressed Quenching and Strong-Coupling of Purcell-Enhanced Single-Molecule Emission in Plasmonic Nanocavities. *ACS Photonics* 5:186-91
54. Guillaume CBal, Bonnot A, Debever JM. 1970. Luminescence from Polaritons. *Physical Review Letters* 24:1235-8
55. Weisbuch C, Nishioka M, Ishikawa A, Arakawa Y. 1992. Observation of the coupled exciton-photon mode splitting in a semiconductor quantum microcavity. *Physical Review Letters* 69:3314-7
56. Khitrova G, Gibbs HM, Jahnke F, Kira M, Koch SW. 1999. Nonlinear optics of normal-mode-coupling semiconductor microcavities. *Reviews of Modern Physics* 71:1591-639
57. Lidzey DG, Bradley DDC, Virgili T, Armitage A, Skolnick MS, Walker S. 1999. Room Temperature Polariton Emission from Strongly Coupled Organic Semiconductor Microcavities. *Physical Review Letters* 82:3316-9
58. Lidzey DG. 2003. Strong Optical Coupling in Organic Semiconductor Microcavities. In *Thin Films and Nanostructures*, ed. VM Agranovich, GF Bassani, 31:355-402: Academic Press. Number of 355-402 pp.
59. Hobson PA, Barnes WL, Lidzey DG, Gehring GA, Whittaker DM, et al. 2002. Strong exciton–photon coupling in a low-Q all-metal mirror microcavity. *Applied Physics Letters* 81:3519-21
60. Schwartz T, Hutchison JA, Genet C, Ebbesen TW. 2011. Reversible Switching of Ultrastrong Light-Molecule Coupling. *Phys. Rev. Lett.* 106:196405

61. Dintinger J, Klein S, Bustos F, Barnes WL, Ebbesen TW. 2005. Strong coupling between surface plasmon-polaritons and organic molecules in subwavelength hole arrays. *Phys. Rev. B* 71:035424
62. Vasa P, Lienau C. 2018. Strong Light–Matter Interaction in Quantum Emitter/Metal Hybrid Nanostructures. *ACS Photonics* 5:2-23
63. Beane G, Brown BS, Johns P, Devkota T, Hartland GV. 2018. Strong Exciton–Plasmon Coupling in Silver Nanowire Nanocavities. *J. Phys. Chem. Lett.* 9:1676-81
64. Bisht A, Cuadra J, Wersäll M, Canales A, Antosiewicz TJ, Shegai T. 2019. Collective Strong Light-Matter Coupling in Hierarchical Microcavity-Plasmon-Exciton Systems. *Nano Letters* 19:189-96
65. Melnikau D, Govyadinov AA, Sánchez-Iglesias A, Grzelczak M, Nabiev IR, et al. 2019. Double Rabi Splitting in a Strongly Coupled System of Core–Shell Au@Ag Nanorods and J-Aggregates of Multiple Fluorophores. *The Journal of Physical Chemistry Letters* 10:6137-43
66. Norris TB, Rhee JK, Sung CY, Arakawa Y, Nishioka M, Weisbuch C. 1994. Time-resolved vacuum Rabi oscillations in a semiconductor quantum microcavity. *Physical Review B* 50:14663-6
67. Brune M, Schmidt-Kaler F, Maali A, Dreyer J, Hagley E, et al. 1996. Quantum Rabi Oscillation: A Direct Test of Field Quantization in a Cavity. *Phys. Rev. Lett.* 76:1800-3
68. Vasa P, Wang W, Pomraenke R, Lammers M, Maiuri M, et al. 2013. Real-time observation of ultrafast Rabi oscillations between excitons and plasmons in metal nanostructures with J-aggregates. *Nat Photon* 7:128-32
69. Vasa P. 2020. Exciton-surface plasmon polariton interactions. *Advances in Physics: X* 5:1749884
70. Kéna-Cohen S, Maier SA, Bradley DDC. 2013. Ultrastrongly Coupled Exciton–Polaritons in Metal-Clad Organic Semiconductor Microcavities. *Adv. Opt. Mater.* 1:827-33
71. Gambino S, Mazzeo M, Genco A, Di Stefano O, Savasta S, et al. 2014. Exploring Light–Matter Interaction Phenomena under Ultrastrong Coupling Regime. *ACS Photonics* 1:1042-8
72. Thomas PA, Tan WJ, Fernandez HA, Barnes WL. 2020. A new signature for strong light-matter coupling using spectroscopic ellipsometry. *Nano Letters*
73. Damari R, Weinberg O, Krotkov D, Demina N, Akulov K, et al. 2019. Strong coupling of collective intermolecular vibrations in organic materials at terahertz frequencies. *Nat. Commun.* 10:3248
74. Imran I, Nicolai GE, Stavinski ND, Sparks JR. 2019. Tuning Vibrational Strong Coupling with Co-Resonators. *ACS Photonics*
75. Takele WM, Wackenhut F, Piatkowski L, Meixner AJ, Waluk J. 2020. Multimode Vibrational Strong Coupling of Methyl Salicylate to a Fabry–Pérot Microcavity. *The Journal of Physical Chemistry B* 124:5709-16
76. Menghrajani KS, Nash GR, Barnes WL. 2019. Vibrational Strong Coupling with Surface Plasmons and the Presence of Surface Plasmon Stop Bands. *ACS Photonics* 6:2110-6
77. Brawley ZT, Storm SD, Contreras Mora DA, Pelton M, Sheldon M. 2021. Angle-independent plasmonic substrates for multi-mode vibrational strong coupling with molecular thin films. *The Journal of Chemical Physics* 154:104305

78. Daehan Yoo FdL-P, In-Ho Lee, Daniel A. Mohr, Matthew Pelton, Markus B. Raschke, Joshua D. Caldwell, Luis Martín-Moreno, Sang-Hyun Oh. 2020. Ultrastrong plasmon-phonon coupling via epsilon-near-zero nanocavities. *arXiv:2003.00136*
79. Dayal G, Morichika I, Ashihara S. 2021. Vibrational Strong Coupling in Subwavelength Nanogap Patch Antenna at the Single Resonator Level. *The Journal of Physical Chemistry Letters* 12:3171-5
80. Zhong X, Chervy T, Wang S, George J, Thomas A, et al. 2016. Non-Radiative Energy Transfer Mediated by Hybrid Light-Matter States. *Angewandte Chemie International Edition* 55:6202-6
81. Zhong X, Chervy T, Zhang L, Thomas A, George J, et al. 2017. Energy Transfer between Spatially Separated Entangled Molecules. *Angewandte Chemie International Edition* 56:9034-8
82. Rozenman GG, Akulov K, Golombek A, Schwartz T. 2018. Long-Range Transport of Organic Exciton-Polaritons Revealed by Ultrafast Microscopy. *ACS Photonics* 5:105-10
83. Xiang B, Ribeiro RF, Du M, Chen L, Yang Z, et al. 2020. Intermolecular vibrational energy transfer enabled by microcavity strong light–matter coupling. *Science* 368:665
84. Georgiou K, Jayaprakash R, Othonos A, Lidzey D. 2021. Ultralong-range polariton-assisted energy transfer in organic microcavities. *Angew Chem Int Ed Engl*
85. Schachenmayer J, Genes C, Tignone E, Pupillo G. 2015. Cavity-Enhanced Transport of Excitons. *Phys. Rev. Lett.* 114:196403
86. Schäfer C, Ruggenthaler M, Appel H, Rubio A. 2019. Modification of excitation and charge transfer in cavity quantum-electrodynamical chemistry. *PNAS* 116:4883-92
87. Aeschlimann M, Brixner T, Cinchetti M, Frisch B, Hecht B, et al. 2017. Cavity-assisted ultrafast long-range periodic energy transfer between plasmonic nanoantennas. *Light: Science & Applications* 6:e17111-e
88. Du M, Martínez-Martínez LA, Ribeiro RF, Hu Z, Menon VM, Yuen-Zhou J. 2018. Theory for polariton-assisted remote energy transfer. *Chem. Sci.* 9:6659-69
89. Rustomji K, Dubois M, Kuhlmey B, de Sterke CM, Enoch S, et al. 2019. Direct Imaging of the Energy-Transfer Enhancement between Two Dipoles in a Photonic Cavity. *Physical Review X* 9:011041
90. Orgiu E, George J, Hutchison JA, Devaux E, Dayen JF, et al. 2015. Conductivity in organic semiconductors hybridized with the vacuum field. *Nat Mater* 14:1123-9
91. Krainova N, Grede AJ, Tsokkou D, Banerji N, Giebink NC. 2020. Polaron Photoconductivity in the Weak and Strong Light-Matter Coupling Regime. *Physical Review Letters* 124:177401
92. Halbhuber M, Mornhinweg J, Zeller V, Ciuti C, Bougeard D, et al. 2020. Non-adiabatic stripping of a cavity field from electrons in the deep-strong coupling regime. *Nature Photonics*
93. Hagenmüller D, Schachenmayer J, Schütz S, Genes C, Pupillo G. 2017. Cavity-Enhanced Transport of Charge. *Phys. Rev. Lett.* 119:223601
94. Hagenmüller D, Schütz S, Schachenmayer J, Genes C, Pupillo G. 2018. Cavity-assisted mesoscopic transport of fermions: Coherent and dissipative dynamics. *Phys. Rev. B* 97:205303

95. Pang Y, Thomas A, Nagarajan K, Vergauwe RMA, Joseph K, et al. 2020. On the Role of Symmetry in Vibrational Strong Coupling: The Case of Charge-Transfer Complexation. *Angewandte Chemie International Edition* 59:10436-40
96. Yang C, Wei X, Sheng J, Wu H. 2020. Phonon heat transport in cavity-mediated optomechanical nanoresonators. *Nat. Commun.* 11:4656
97. Persson BNJ, Kato T, Ueba H, Volokitin AI. 2007. Vibrational heating of molecules adsorbed on insulating surfaces using localized photon tunneling. *Phys. Rev. B* 75
98. Kim K, Song B, Fernández-Hurtado V, Lee W, Jeong W, et al. 2015. Radiative heat transfer in the extreme near field. *Nature* 528:387-91
99. Delor M, Scattergood PA, Sazanovich IV, Parker AW, Greetham GM, et al. 2014. Toward control of electron transfer in donor-acceptor molecules by bond-specific infrared excitation. *Science* 346:1492-5
100. Delor M, Keane T, Scattergood PA, Sazanovich IV, Greetham GM, et al. 2015. On the mechanism of vibrational control of light-induced charge transfer in donor-bridge-acceptor assemblies. *Nat. Chem.* 7:689-95
101. Sukegawa J, Schubert C, Zhu X, Tsuji H, Guldi DM, Nakamura E. 2014. Electron transfer through rigid organic molecular wires enhanced by electronic and electron–vibration coupling. *Nat. Chem.* 6:899-905
102. Skourtis SS, Waldeck DH, Beratan DN. 2004. Inelastic electron tunneling erases coupling-pathway interferences. *J. Phys. Chem. B* 108:15511-8
103. Carias H, Beratan DN, Skourtis SS. 2011. Floquet Analysis for Vibronically Modulated Electron Tunneling. *J. Phys. Chem. B* 115:5510-8
104. Ballmann S, Härtle R, Coto PB, Elbing M, Mayor M, et al. 2012. Experimental Evidence for Quantum Interference and Vibrationally Induced Decoherence in Single-Molecule Junctions. *Phys. Rev. Lett.* 109:056801
105. Romero E, Augulis R, Novoderezhkin VI, Ferretti M, Thieme J, et al. 2014. Quantum coherence in photosynthesis for efficient solar-energy conversion. *Nat Phys* 10:676-82
106. Wang T, Kafle TR, Kattel B, Chan W-L. 2016. Observation of an Ultrafast Exciton Hopping Channel in Organic Semiconducting Crystals. *J. Phys. Chem. C* 120:7491-9
107. Chin AW, Prior J, Rosenbach R, Caycedo-Soler F, Huelga SF, Plenio MB. 2013. The role of non-equilibrium vibrational structures in electronic coherence and recoherence in pigment–protein complexes. *Nature Physics* 9:113-8
108. Tiwari V, Peters WK, Jonas DM. 2013. Electronic resonance with anticorrelated pigment vibrations drives photosynthetic energy transfer outside the adiabatic framework. *Proceedings of the National Academy of Sciences* 110:1203
109. O'Reilly EJ, Olaya-Castro A. 2014. Non-classicality of the molecular vibrations assisting exciton energy transfer at room temperature. *Nat. Commun.* 5:3012
110. Nelson TR, Ondarse-Alvarez D, Oldani N, Rodriguez-Hernandez B, Alfonso-Hernandez L, et al. 2018. Coherent exciton-vibrational dynamics and energy transfer in conjugated organics. *Nat. Commun.* 9:2316
111. Goldberg O, Meir Y, Dubi Y. 2018. Vibration-Assisted and Vibration-Hampered Excitonic Quantum Transport. *J. Phys. Chem. Lett.* 9:3143-8
112. Abramavicius D, Valkunas L. 2016. Role of coherent vibrations in energy transfer and conversion in photosynthetic pigment–protein complexes. *Photosynthesis Research* 127:33-47

113. Hua XM, Gersten JI, Nitzan A. 1985. Theory of energy transfer between molecules near solid-dtate particles. *J. Chem. Phys.* 83:3650-9
114. Pustovit VN, Shahbazyan TV. 2011. Resonance energy transfer near metal nanostructures mediated by surface plasmons. *Phys. Rev. B* 83:085427
115. Pustovit VN, Urbas AM, Shahbazyan TV. 2014. Energy transfer in plasmonic systems. *Journal of Optics* 16:114015
116. Hsu L-Y, Ding W, Schatz GC. 2017. Plasmon-Coupled Resonance Energy Transfer. *J. Phys. Chem. Lett.* 8:2357-67
117. Wu JS, Lin YC, Shen YL, Hsu LY. 2018. Characteristic Distance of Resonance Energy Transfer Coupled with Surface Plasmon Polaritons. *J. Phys. Chem. Lett.* 9:7032-9
118. Hutchison JA, Liscio A, Schwartz T, Canaguier-Durand A, Genet C, et al. 2013. Tuning the Work-Function Via Strong Coupling. *Advanced Materials* 25:2481-5
119. Wang K, Seidel M, Nagarajan K, Chervy T, Genet C, Ebbesen T. 2021. Large optical nonlinearity enhancement under electronic strong coupling. *Nat. Commun.* 12:1486
120. Hirai K, Takeda R, Hutchison JA, Uji-i H. 2020. Modulation of Prins Cyclization by Vibrational Strong Coupling. *Angewandte Chemie International Edition* 59:5332-5
121. Hirai K, Ishikawa H, Chervy T, HUTCHISON J, Uji-i H. 2021. Selective Crystallization via Vibrational Strong Coupling. *chemrxiv*
122. Pietron JJ, Fears KP, Owrutsky JC, Simpkins BS. 2020. Electrochemical Modulation of Strong Vibration–Cavity Coupling. *ACS Photonics* 7:165-73
123. Xiang B, Ribeiro RF, Li Y, Dunkelberger AD, Simpkins BB, et al. 2019. Manipulating optical nonlinearities of molecular polaritons by delocalization. *Sci. Adv.* 5:eaax5196
124. Xiang B, Ribeiro RF, Chen L, Wang J, Du M, et al. 2019. State-Selective Polariton to Dark State Relaxation Dynamics. *The Journal of Physical Chemistry A* 123:5918-27
125. Ruggenthaler M, Flick J, Pellegrini C, Appel H, Tokatly IV, Rubio A. 2014. Quantum-electrodynamical density-functional theory: Bridging quantum optics and electronic-structure theory. *Physical Review A* 90:012508
126. Flick J, Ruggenthaler M, Appel H, Rubio A. 2015. Kohn–Sham approach to quantum electrodynamical density-functional theory: Exact time-dependent effective potentials in real space. *PNAS* 112:15285-90
127. Pellegrini C, Flick J, Tokatly IV, Appel H, Rubio A. 2015. Optimized Effective Potential for Quantum Electrodynamical Time-Dependent Density Functional Theory. *Phys. Rev. Lett.* 115:093001
128. Flick J, Ruggenthaler M, Appel H, Rubio A. 2017. Atoms and molecules in cavities, from weak to strong coupling in quantum-electrodynamics (QED) chemistry. *PNAS* 114:3026-34
129. Schäfer C, Ruggenthaler M, Rubio A. 2018. Ab initio nonrelativistic quantum electrodynamics: Bridging quantum chemistry and quantum optics from weak to strong coupling. *Phys. Rev. A* 98:043801
130. Flick J, Narang P. 2018. Cavity-Correlated Electron-Nuclear Dynamics from First Principles. *Physical Review Letters* 121:113002
131. Flick J, Welakuh DM, Ruggenthaler M, Appel H, Rubio A. 2019. Light-Matter Response in Non-Relativistic Quantum Electrodynamics. *ACS Photonics*
132. Lacombe L, Hoffmann NM, Maitra NT. 2019. Exact Potential Energy Surface for Molecules in Cavities. *arXiv:1906.02651v1*

133. Haugland TS, Ronca E, Kjønstad EF, Rubio A, Koch H. 2020. Coupled Cluster Theory for Molecular Polaritons: Changing Ground and Excited States. *Physical Review X* 10:041043
134. Hoffmann NM, Lacombe L, Rubio A, Maitra NT. 2020. Effect of many modes on self-polarization and photochemical suppression in cavities. *The Journal of Chemical Physics* 153:104103
135. Flick J, Narang P. 2020. Ab initio polaritonic potential-energy surfaces for excited-state nanophotonics and polaritonic chemistry. *The Journal of Chemical Physics* 153:094116
136. Haugland TS, Schäfer C, Ronca E, Rubio A, Koch H. 2021. Intermolecular interactions in optical cavities: An ab initio QED study. *The Journal of Chemical Physics* 154:094113
137. Schäfer C, Flick J, Ronca E, Narang P, Rubio A. 2021. Shining Light on the Microscopic Resonant Mechanism Responsible for Cavity-Mediated Chemical Reactivity.
138. Gersten JI, Nitzan A. 1985. Photophysics and photochemistry near surfaces and small particles. *Surf. Sci.* 158:165-89
139. Clavero C. 2014. Plasmon-induced hot-electron generation at nanoparticle/metal-oxide interfaces for photovoltaic and photocatalytic devices. *Nature Photonics* 8:95-103
140. Shan H, Yu Y, Wang X, Luo Y, Zu S, et al. 2019. Direct observation of ultrafast plasmonic hot electron transfer in the strong coupling regime. *Light: Science & Applications* 8:9
141. Luk HL, Feist J, Toppari JJ, Groenhof G. 2017. Multiscale Molecular Dynamics Simulations of Polaritonic Chemistry. *Journal of Chemical Theory and Computation* 13:4324-35
142. Groenhof G, Climent C, Feist J, Morozov D, Toppari JJ. 2019. Tracking Polariton Relaxation with Multiscale Molecular Dynamics Simulations. *The Journal of Physical Chemistry Letters* 10:5476-83
143. Tichauer RH, Feist J, Groenhof G. 2021. Multi-scale dynamics simulations of molecular polaritons: The effect of multiple cavity modes on polariton relaxation. *The Journal of Chemical Physics* 154:104112
144. Hutchison JA, Schwartz T, Genet C, Devaux E, Ebbesen TW. 2012. Modifying Chemical Landscapes by Coupling to Vacuum Fields. *Angew. Chem. Int. Ed.* 51:1592-6
145. Peters VN, Faruk MO, Asane J, Alexander R, Peters DaA, et al. 2019. Effect of strong coupling on photodegradation of the semiconducting polymer P3HT. *Optica* 6:318-25
146. Munkhbat B, Wersäll M, Baranov DG, Antosiewicz TJ, Shegai T. 2018. Suppression of photo-oxidation of organic chromophores by strong coupling to plasmonic nanoantennas. *Sci. Adv.* 4:eaas9552
147. Galego J, Garcia-Vidal FJ, Feist J. 2016. Suppressing photochemical reactions with quantized light fields. *Nat. Commun.* 7:13841
148. Mandal A, Huo P. 2019. Investigating New Reactivities Enabled by Polariton Photochemistry. *The Journal of Physical Chemistry Letters* 10:5519-29
149. Mandal A, Krauss TD, Huo PF. 2020. Polariton-Mediated Electron Transfer via Cavity Quantum Electrodynamics. *Journal of Physical Chemistry B* 124:6321-40

150. Herrera F, Spano FC. 2016. Cavity-Controlled Chemistry in Molecular Ensembles. *Phys. Rev. Lett.* 116:238301
151. Galego J, Garcia-Vidal FJ, Feist J. 2017. Many-Molecule Reaction Triggered by a Single Photon in Polaritonic Chemistry. *Phys. Rev. Lett.* 119:136001
152. Mauro L, Caicedo K, Jonusauskas G, Avriller R. 2021. Charge-transfer chemical reactions in nanofluidic Fabry-P\'erot cavities. *Physical Review B* 103:165412
153. Thomas A, George J, Shalabney A, Dryzhakov M, Varma SJ, et al. 2016. Ground-State Chemical Reactivity under Vibrational Coupling to the Vacuum Electromagnetic Field. *Angew. Chem. Int. Ed.* 55:11462-6
154. Thomas A, Lethuillier-Karl L, Nagarajan K, Vergauwe RMA, George J, et al. 2019. Tilting a ground-state reactivity landscape by vibrational strong coupling. *Science* 363:615
155. Lather J, Bhatt P, Thomas A, Ebbesen TW, George J. 2019. Cavity Catalysis by Cooperative Vibrational Strong Coupling of Reactant and Solvent Molecules. *Angew. Chem. Int. Ed.* 58:10635-8
156. Vergauwe RMA, Thomas A, Nagarajan K, Shalabney A, George J, et al. 2019. Modification of Enzyme Activity by Vibrational Strong Coupling of Water. *Angewandte Chemie International Edition* 58:15324-8
157. Thomas A, Jayachandran A, Lethuillier-Karl L, Vergauwe RMA, Nagarajan K, et al. 2020. Ground state chemistry under vibrational strong coupling: dependence of thermodynamic parameters on the Rabi splitting energy. *Nanophotonics* 9:249-55
158. Sau A, Nagarajan K, Patrahau B, Lethuillier-Karl L, Vergauwe R, et al. 2020. Modifying Woodward-Hoffmann Stereoselectivity under Vibrational Strong Coupling. *Angewandte Chemie International Edition* n/a
159. Lather J, George J. 2021. Improving Enzyme Catalytic Efficiency by Co-operative Vibrational Strong Coupling of Water. *J. Phys. Chem. Lett.* 12:379-84
160. Hidefumi H, Atef S, Jino G. 2019. Vacuum-Field Catalysis: Accelerated Reactions by Vibrational Ultra Strong Coupling. *10.26434/chemrxiv.7234721.v4*
161. Imperatore MV, Asbury JB, Giebink NC. 2021. Reproducibility of cavity-enhanced chemical reaction rates in the vibrational strong coupling regime. *The Journal of Chemical Physics* 154:191103
162. Hirai K, Hutchison JA, Uji-i H. 2020. Recent Progress in Vibropolaritonic Chemistry. *ChemPlusChem* 85:1981-8
163. Dunkelberger AD, Spann BT, Fears KP, Simpkins BS, Owrutsky JC. 2016. Modified relaxation dynamics and coherent energy exchange in coupled vibration-cavity polaritons. *Nat. Commun.* 7:13504
164. Galego J, Garcia-Vidal FJ, Feist J. 2015. Cavity-Induced Modifications of Molecular Structure in the Strong-Coupling Regime. *Phys. Rev. X* 5:041022
165. Wu N, Feist J, Garcia-Vidal FJ. 2016. When polarons meet polaritons: Exciton-vibration interactions in organic molecules strongly coupled to confined light fields. *Physical Review B* 94:195409
166. Spano FC. 2015. Optical microcavities enhance the exciton coherence length and eliminate vibronic coupling in J-aggregates. *The J. Chem. Phys.* 142:184707
167. Zeb MA, Kirton PG, Keeling J. 2018. Exact States and Spectra of Vibrationally Dressed Polaritons. *ACS Photonics* 5:249-57

168. Holstein T. 1959. Studies of polaron motion: Part I. The molecular-crystal model. *Annals of Physics* 8:325-42
169. Holstein T. 1959. Studies of polaron motion: Part II. The "small" polaron. *Annals of Physics* 8:343-89
170. Spano FC. 2010. The Spectral Signatures of Frenkel Polarons in H- and J-Aggregates. *Accounts of Chemical Research* 43:429-39
171. Herrera F, Spano FC. 2017. Dark Vibronic Polaritons and the Spectroscopy of Organic Microcavities. *Phys. Rev. Lett.* 118:223601
172. Herrera F, Spano FC. 2017. Absorption and photoluminescence in organic cavity QED. *Phys. Rev. A* 95:053867
173. Avramenko AG, Rury AS. 2020. Quantum Control of Ultrafast Internal Conversion Using Nanoconfined Virtual Photons. *The Journal of Physical Chemistry Letters* 11:1013-21
174. Senitzky IR. 1960. Induced and Spontaneous Emission in a Coherent Field. III. *Phys. Rev.* 119:1807-15
175. Senitzky IR. 1961. Onset of Correlation in Initially Uncorrelated System. *Phys. Rev.* 121:171-81
176. Shahbazyan TV, Raikh ME, Vardeny ZV. 2000. Mesoscopic cooperative emission from a disordered system. *Phys. Rev. B* 61:13266-76
177. Celardo GL, Biella A, Kaplan L, Borgonovi F. 2013. Interplay of superradiance and disorder in the Anderson Model. *Fortschritte der Physik* 61:250-60
178. Celardo GL, Giusteri GG, Borgonovi F. 2014. Cooperative robustness to static disorder: Superradiance and localization in a nanoscale ring to model light-harvesting systems found in nature. *Phys. Rev. B* 90:075113
179. Sun C, Chernyak VY, Piryatinski A, Sinitsyn NA. 2019. Cooperative Light Emission in the Presence of Strong Inhomogeneous Broadening. *Physical Review Letters* 123:123605
180. Manceau JM, Biasiol G, Tran NL, Carusotto I, Colombelli R. 2017. Immunity of intersubband polaritons to inhomogeneous broadening. *Physical Review B* 96:235301
181. Houdré R, Stanley RP, Ilegems M. 1996. Vacuum-field Rabi splitting in the presence of inhomogeneous broadening: Resolution of a homogeneous linewidth in an inhomogeneously broadened system. *Physical Review A* 53:2711-5
182. Nitzan A. 2006. *Chemical Dynamics in Condensed Phases*. Oxford: Oxford Univ. Press
183. Galego J, Climent C, Garcia-Vidal FJ, Feist J. 2019. Cavity Casimir-Polder Forces and Their Effects in Ground-State Chemical Reactivity. *Physical Review X* 9:021057
184. Li TE, Nitzan A, Subotnik JE. 2020. On the origin of ground-state vacuum-field catalysis: Equilibrium consideration. *The Journal of Chemical Physics* 152:234107
185. Campos-Gonzalez-Angulo JA, Yuen-Zhou J. 2020. Polaritonic normal modes in transition state theory. *The Journal of Chemical Physics* 152:161101
186. Zhdanov VP. 2020. Vacuum field in a cavity, light-mediated vibrational coupling, and chemical reactivity. *Chemical Physics*:110767
187. Fischer EW, Saalfrank P. 2021. Ground state properties and infrared spectra of anharmonic vibrational polaritons of small molecules in cavities. *The Journal of Chemical Physics* 154:104311

188. Climent C, Feist J. 2020. On the SN2 reactions modified in vibrational strong coupling experiments: reaction mechanisms and vibrational mode assignments. *Physical Chemistry Chemical Physics*
189. Triana F, Herrera F. 2020. Self-dissociation of polar molecules in a confined infrared vacuumJohan. *chemrxiv*
190. Li X, Mandal A, Huo P. 2021. Cavity frequency-dependent theory for vibrational polariton chemistry. *Nat. Commun.* 12:1315
191. Semenov A, Nitzan A. 2019. Electron transfer in confined electromagnetic fields. *J. Chem. Phys.* 150:174122
192. Phuc NT, Trung PQ, Ishizaki A. 2020. Controlling the nonadiabatic electron-transfer reaction rate through molecular-vibration polaritons in the ultrastrong coupling regime. *Sci Rep* 10:7318
193. Campos-Gonzalez-Angulo JA, Ribeiro RF, Yuen-Zhou J. 2019. Resonant catalysis of thermally activated chemical reactions with vibrational polaritons. *Nat. Commun.* 10:4685
194. Vurgaftman I, Simpkins BS, Dunkelberger AD, Owrutsky JC. 2020. Negligible Effect of Vibrational Polaritons on Chemical Reaction Rates via the Density of States Pathway. *The Journal of Physical Chemistry Letters*
195. Wellnitz D, Schütz S, Whitlock S, Schachenmayer J, Pupillo G. 2020. Collective Dissipative Molecule Formation in a Cavity. *Physical Review Letters* 125:193201
196. Saurabh P, Mukamel S. 2016. Two-dimensional infrared spectroscopy of vibrational polaritons of molecules in an optical cavity. *The Journal of Chemical Physics* 144:124115
197. F. Ribeiro R, Dunkelberger AD, Xiang B, Xiong W, Simpkins BS, et al. 2018. Theory for Nonlinear Spectroscopy of Vibrational Polaritons. *J. Phys. Chem. Lett.* 9:3766-71
198. Hernández FJ, Herrera F. 2019. Multi-level quantum Rabi model for anharmonic vibrational polaritons. *The Journal of Chemical Physics* 151:144116
199. Pino Jd, Feist J, Garcia-Vidal FJ. 2015. Quantum theory of collective strong coupling of molecular vibrations with a microcavity mode. *New J. Phys.* 17:053040
200. Gonzalez-Ballestero C, Feist J, Gonzalo Badía E, Moreno E, Garcia-Vidal FJ. 2016. Uncoupled Dark States Can Inherit Polaritonic Properties. *Physical Review Letters* 117:156402
201. Li TE, Nitzan A, Subotnik JE. 2021. Collective vibrational strong coupling effects on molecular vibrational relaxation and energy transfer: Numerical insights via cavity molecular dynamics simulations. *Angewandte Chemie International Edition* n/a
202. Du M, Yuen-Zhou J. 2021. Can Dark States Explain Vibropolaritonic Chemistry? *arXiv:2104.07214v1 [quant-ph]*
203. Sidler D, Schäfer C, Ruggenthaler M, Rubio A. 2021. Polaritonic Chemistry: Collective Strong Coupling Implies Strong Local Modification of Chemical Properties. *The Journal of Physical Chemistry Letters* 12:508-16
204. Li TE, Nitzan A, Subotnik JE. 2021. Energy-efficient pathway for selectively exciting solute molecules to high vibrational states via solvent vibration-polariton pumping. *arXiv:2104.15121*
205. Davidsson E, Kowalewski M. 2020. Atom Assisted Photochemistry in Optical Cavities. *The Journal of Physical Chemistry A* 124:4672-7

206. Groenhof G, Toppari JJ. 2018. Coherent Light Harvesting through Strong Coupling to Confined Light. *J. Phys. Chem. Lett.* 9:4848-51
207. Li TE, Subotnik JE, Nitzan A. 2020. Cavity molecular dynamics simulations of liquid water under vibrational ultrastrong coupling. *Proc. Nat. Acad. Sci.* 117:18324
208. Li TE, Nitzan A, Subotnik JE. 2021. Cavity molecular dynamics simulations of vibrational polariton-enhanced molecular nonlinear absorption. *The Journal of Chemical Physics* 154:094124
209. Shalabney A, George J, Hiura H, Hutchison JA, Genet C, et al. 2015. Enhanced Raman Scattering from Vibro-Polariton Hybrid States. *Angewandte Chemie International Edition* 54:7971-5
210. del Pino J, Feist J, Garcia-Vidal FJ. 2015. Signatures of Vibrational Strong Coupling in Raman Scattering. *The Journal of Physical Chemistry C* 119:29132-7
211. Ahn W, Simpkins BS. 2020. Raman Scattering under Strong Vibration-Cavity Coupling. *The Journal of Physical Chemistry C*
212. Baieva S, Ihalainen JA, Toppari JJ. 2013. Strong coupling between surface plasmon polaritons and β-carotene in nanolayered system. *The Journal of Chemical Physics* 138:044707
213. Nagasawa F, Takase M, Murakoshi K. 2014. Raman Enhancement via Polariton States Produced by Strong Coupling between a Localized Surface Plasmon and Dye Excitons at Metal Nanogaps. *The Journal of Physical Chemistry Letters* 5:14-9
214. Avramenko AG, Rury AS. 2019. Interrogating the Structure of Molecular Cavity Polaritons with Resonance Raman Scattering: An Experimentally Motivated Theoretical Description. *The Journal of Physical Chemistry C* 123:30551-61
215. Neuman T, Aizpurua J, Esteban R. 2020. Quantum theory of surface-enhanced resonant Raman scattering (SERRS) of molecules in strongly coupled plasmon-exciton systems. *Nanophotonics* 9:295-308
216. Golombek A, Balasubrahmaniyam M, Kaeek M, Hadar K, Schwartz T. 2020. Collective Rayleigh Scattering from Molecular Ensembles under Strong Coupling. *The Journal of Physical Chemistry Letters* 11:3803-8
217. Aspelmeyer M, Kippenberg TJ, Marquardt F. 2014. Cavity optomechanics. *Rev. Mod. Phys.* 86:1391-452
218. Olaimat MM, Yousefi L, Ramahi OM. 2021. Using plasmonics and nanoparticles to enhance the efficiency of solar cells: review of latest technologies. *J. Opt. Soc. Am. B* 38:638-51
219. Coccia E, Fregoni J, Guido CA, Marsili M, Pipolo S, Corni S. 2020. Hybrid theoretical models for molecular nanoplasmonics. *The Journal of Chemical Physics* 153:200901
220. Martínez-Martínez LA, Du M, Ribeiro RF, Kéna-Cohen S, Yuen-Zhou J. 2018. Polariton-Assisted Singlet Fission in Acene Aggregates. *J. Phys. Chem. Lett.* 9:1951-7
221. Martínez-Martínez LA, Eizner E, Kéna-Cohen S, Yuen-Zhou J. 2019. Triplet harvesting in the polaritonic regime: A variational polaron approach. *The Journal of Chemical Physics* 151:054106
222. Polak D, Jayaprakash R, Lyons TP, Martínez-Martínez LÁ, Leventis A, et al. 2020. Manipulating molecules with strong coupling: harvesting triplet excitons in organic exciton microcavities. *Chem. Sci.* 11:343-54

223. Ye C, Mallick S, Hertzog M, Kowalewski M, Börjesson K. 2021. Direct Transition from Triplet Excitons to Hybrid Light–Matter States via Triplet–Triplet Annihilation. *Journal of the American Chemical Society* 143:7501-8
224. Bera ML, Julià-Farré S, Lewenstein M, Bera MN. 2021. Quantum Heat Engines with Carnot Efficiency at Maximum Power. *arXiv:2106.01193v1*
225. Campaioli F, Pollock FA, Vinjanampathy S. 2018. Quantum Batteries - Review Chapter. *arXiv:1805.05507 [quant-ph]*
226. Alicki R, Fannes M. 2013. Entanglement boost for extractable work from ensembles of quantum batteries. *Phys. Rev. E* 87:042123
227. Hovhannisyan KV, Perarnau-Llobet M, Huber M, Acín A. 2013. Entanglement Generation is Not Necessary for Optimal Work Extraction. *Phys. Rev. Lett.* 111:240401
228. Campaioli F, Pollock FA, Binder FC, Céleri L, Goold J, et al. 2017. Enhancing the Charging Power of Quantum Batteries. *Phys. Rev. Lett.* 118:150601
229. Le TP, Levinsen J, Modi K, Parish MM, Pollock FA. 2018. Spin-chain model of a many-body quantum battery. *Phys. Rev. A* 97:022106
230. Ferraro D, Campisi M, Andolina GM, Pellegrini V, Polini M. 2018. High-Power Collective Charging of a Solid-State Quantum Battery. *Phys. Rev. Lett.* 120:117702
231. Andolina GM, Keck M, Mari A, Campisi M, Giovannetti V, Polini M. 2019. Extractable Work, the Role of Correlations, and Asymptotic Freedom in Quantum Batteries. *Phys. Rev. Lett.* 122:047702
232. Pirmoradian F, Mølmer K. 2019. Aging of a quantum battery. *Physical Review A* 100:043833
233. Tabesh FT, Kamin FH, Salimi S. 2020. Environment-mediated charging process of quantum batteries. *Physical Review A* 102:052223
234. Santos AC. 2021. Quantum advantage of two-level batteries in the self-discharging process. *Physical Review E* 103:042118
235. Niedenzu W, Kurizki G. 2018. Cooperative many-body enhancement of quantum thermal machine power. *New J. Phys.* 20:113038
236. Kloc M, Cejnar P, Schaller G. 2019. Collective performance of a finite-time quantum Otto cycle. *Physical Review E* 100:042126
237. Xiang B, Wang J, Yang Z, Xiong W. 2021. Nonlinear infrared polaritonic interaction between cavities mediated by molecular vibrations at ultrafast time scale. *Sci. Adv.* 7:eabf6397
238. Stensitzki T, Yang Y, Kozich V, Ahmed AA, Kössl F, et al. 2018. Acceleration of a ground-state reaction by selective femtosecond-infrared-laser-pulse excitation. *Nat. Chem.* 10:126-31
239. Heyne K, Kühn O. 2019. Infrared Laser Excitation Controlled Reaction Acceleration in the Electronic Ground State. *Journal of the American Chemical Society* 141:11730-8
240. Velez ST, Pogrebna A, Galland C. 2021. Collective Vibrational Quantum Coherence in a Molecular Liquid under Spontaneous Raman Scattering. *arXiv:2105.00213 [quant-ph]*